\documentclass[fleqn,usenatbib]{mnras}
\usepackage{newtxtext,newtxmath}
\usepackage[T1]{fontenc}
\usepackage{ae,aecompl}

\usepackage{graphicx}	% Including figure files
\usepackage{amsmath}	% Advanced maths commands
\usepackage{amssymb}	% Extra maths symbols
\usepackage{pdflscape}  % to produce a landscape table
\title[{\it Gaia} Nuclear Transients]{{\it Gaia} transients in galactic nuclei}

\author[Z. Kostrzewa-Rutkowska et al.]{
Z. Kostrzewa-Rutkowska$^{1,2}$\thanks{E-mail: z.p.kostrzewa@sron.nl},
P.G. Jonker$^{1,2}$, S.T. Hodgkin$^{3}$,
{\L}. Wyrzykowski$^{4}$,\newauthor
M. Fraser$^{5}$, D.L. Harrison$^{3,6}$, G. Rixon$^{3}$, A. Yoldas$^{3}$, F. van Leeuwen$^{3}$, \newauthor  
A. Delgado$^{3}$,  M. van Leeuwen$^{3}$, S. E. Koposov$^{7,3}$
\\
$^{1}$SRON, Netherlands Institute for Space Research, Sorbonnelaan 2, 3584 CA Utrecht, the Netherlands\\
$^{2}$Department of Astrophysics/IMAPP, Radboud University, P.O. Box 9010, 6500 GL Nijmegen, the Netherlands\\
$^{3}$Institute of Astronomy, University of Cambridge, Madingley Road, Cambridge CB3 0HA, United Kingdom\\
$^{4}$Warsaw University Astronomical Observatory, Al. Ujazdowskie 4, 00-478 Warszawa, Poland\\
$^{5}$School of Physics, O'Brien Centre for Science North, University College Dublin, Belfield, Dublin 4, Ireland\\
$^{6}$Kavli Institute for Cosmology, University of Cambridge, Madingley Road, Cambridge CB3 0HA, United Kingdom\\
$^{7}$McWilliams Center for Cosmology, Carnegie Mellon University, 5000 Forbes Ave, 15213, USA
}

\date{Accepted XXX. Received YYY; in original form ZZZ}

\pubyear{2018}

\begin{document}
\label{firstpage}
\pagerange{\pageref{firstpage}--\pageref{lastpage}}
\maketitle

% Abstract of the paper
\begin{abstract}

The high spatial resolution and precise astrometry and photometry of the {\it Gaia} mission should make it particularly apt at discovering and resolving transients occurring in, or near, the centres of galaxies. Indeed, some nuclear transients are reported by the {\it Gaia} Science Alerts (GSA) team, but not a single confirmed tidal disruption event (TDE) has been published. In order to explore the sensitivity of GSA, we performed an independent and systematic search for nuclear transients using {\it Gaia} observations.
Our transient search is driven from an input galaxy catalogue (derived from the Sloan Digital Sky Survey Release 12). We present a candidate detection metric that is independent from the existing GSA methodology, to see if {\it Gaia} Alerts are biased against the discovery of nuclear transients, and in particular which steps may have an impact. Our technique does require significant manual vetting of candidates, making implementation in the GSA system impractical for daily operations, although it could be run weekly, which for month-to-year long transients would make a scientifically valuable addition.  
Our search yielded $\sim$480 nuclear transients, 5 of which were alerted and published by GSA. The list of (in some cases ongoing) transients includes candidates for events related to enhanced accretion on to a super-massive black hole and TDEs. An implementation of the detection methodology and criteria used in this paper as an extension of GSA could open up the possibility for {\it Gaia} to fulfil the role as a main tool to find transient nuclear activity as predicted in the literature.

\end{abstract}

%482

% Select between one and six entries from the list of approved keywords.
% Don't make up new ones.
\begin{keywords}
surveys -- galaxies: nuclei --  supernovae: general
\end{keywords}

%%%%%%%%%%%%%%%%%%%%%%%%%%%%%%%%%%%%%%%%%%%%%%%%%%

%%%%%%%%%%%%%%%%% BODY OF PAPER %%%%%%%%%%%%%%%%%%

\section{Introduction}

The European Space Agency (ESA)--{\it Gaia} mission has been operational since mid-2014 and has provided accurate photometric, astrometric, and spectroscopic measurements for roughly a billion stars in the Milky Way (\citealt{2016A&A...595A...1G,2016A&A...595A...2G}). {\it Gaia}'s on-board detection algorithms are optimised for the detection of point-like sources, which implies that extended sources have to have an effective radius less than 0.6 arcsec in order to be detected (\citealt{2014sf2a.conf..421D,2015A&A...576A..74D}). Thus, the mission is also collecting data for a significant number of resolved extragalactic objects, such as small elliptical galaxies, and galaxies with compact bulges, or point-like sources  such as high redshift quasars.
The observing strategy (Nominal Scanning Law) of {\it Gaia} is optimised to deliver data for parallax measurement. As a result of this scanning law, most of the sources will be scanned more than about 70 times from different angles during five year mission. Each position on the sky is scanned, on average, once every 30 days (\citealt{2016A&A...595A...4L}). 

%GaiaAlerts AlertPipe
These repeat visits make {\it Gaia} an all-sky, multi-epoch photometric survey, that allows us to monitor variability with high precision as well as detect new transient sources (\citealt{2013RSPTA.37120239H,2017arXiv170203295E}). The Data Processing and Analysis Consortium (DPAC) {\it Gaia} data flow enables detections of transients within 24-48 hours of the observation (in the best case). However, due to different reasons the delay might be up to a few days (Hodgkin et al. in prep). From September 2014 onwards new transients from {\it Gaia} have been made publicly available after  manual vetting of candidate transients detected by the {\it Gaia} Science Alerts (GSA) team. To this end, AlertPipe - dedicated software for data processing, transient searching, and candidate filtering was employed (Hodgkin et al.~in prep). Such a near-real-time survey is predicted to detect about 6000 low-redshift supernovae brighter than $G=19$ mag and 1300 microlensing events during the first five years of the mission (\citealt{2002MNRAS.331..649B,2003MNRAS.341..569B,2012Ap&SS.341..163A}). Accurate photometry and low--resolution spectroscopy should allow for a robust classification and reduce the rate of false positives. {\it Gaia} could therefore play an important role in transient detection. AlertPipe employs two different transient detection algorithms. Transient discovery is either based on the detection of a new source ({\it NewSource} detector; the event has to have 2 or more observations above flux threshold, equivalent to $G=19$), or a significant deviation in brightness of a known source compared to previous {\it Gaia} photometry ({\it OldSource} detector, either a source brightened by more than 1 magnitude and this rise is more than 3 sigma above the rms of the historic variations from all data available, or a source brightened by more than 0.15 magnitudes and this rise is more than 6 times the rms of the historic variations from all data available). The thresholds and other detection parameters of both detectors are tuneable but have been kept fixed over the period June 2016 -- June 2017 under consideration here.

%nuclears
Optical variability occurring in the centres of galaxies can be associated with transients such as (superluminous) supernovae (SNe) and tidal disruption events (TDEs). On the other hand active galactic nuclei (AGNs) exhibit variability across the whole electromagnetic spectrum caused by activity in the accretion disk and jet (see e.g.~\citealt{2012ApJ...753..106M,2017MNRAS.470.4112G} for recent work showing transient and variability phenomena associated with the central super-massive black hole in galaxies).

Core-Collapse Supernovae (CCSNe) originate from explosions of massive stars (masses $M>8 M_\odot$). CCSNe have been proposed as environmental metallicity
probes (\citealt{2014MNRAS.440.1856D}). Together with Type Ia Supernovae (SNe Ia), they shape and influence galaxy structure and star formation (\citealt{2017ApJ...848...25M}). 
Being standard candles, SNe Ia can also be used as probes of distribution of dust in their host galaxies. Tracing it is particularly important in the very cores of galaxies, typically containing large amounts of obscuring dust, to test the relations between SN Ia observed brightness and distance from the core of their host galaxy as well as morphology of the host.
Superluminous supernovae (SLSNe) are associated with deaths of the most massive stars, which means that they may have an impact on the chemical evolution and re-ionization of the Universe (\citealt{2010ApJ...724L..16P,2012Sci...337..927G}). The SLSN explosions are probably induced by different physical mechanisms than other, more common types of SNe (\citealt{2018MNRAS.475.1046I}). 
Tidal disruption events can be used to determine the presence and study the properties, such as the mass, of supermassive black holes (SMBHs) in quiescent galaxies (\citealt{1988Natur.333..523R}). TDE properties probe the stellar populations and dynamics in galactic nuclei, the physics of black hole accretion including the potential to detect relativistic effects near the SMBH, and the physics of jet formation and evolution (e.g. \citealt{2016MNRAS.461..371K,2018ApJ...852...72V}). In addition, because the rate of TDEs is temporarily massively enhanced in binary SMBH systems, TDEs might point us to galaxies that host compact binary SMBHs (\citealt{2011ApJ...738L...8W,2011ApJ...729...13C}). Finally, the volumetric TDE rate is a proxy for the mass of the black hole seeds that grow into SMBHs (\citealt{2016MNRAS.455..859S}). However, the inhomogeneous and small sample of the events currently available (about several tens\footnote{\tt http://tde.space}) probably prevents us from reaping the full potential of TDE studies. 

Most of the ground based surveys hunting for supernovae as well as spectroscopic follow-up observations had the tendency to avoid the central regions of host galaxies. This was largely due to various difficulties in the data processing and lower signal-to-noise of observed transients due to the core brightness. Although recent developments in difference image analysis techniques mitigate these issues (e.g. \citealt{2016ApJ...830...27Z}), the high spatial resolution afforded by {\it Gaia} and the lack of atmospheric seeing variations should also allow {\it Gaia} to resolve transients at closer angular separations to their host galaxy nuclei and enable discrimination between genuinely nuclear transients  (e.g. TDEs) and near-nuclear events (e.g. circumnuclear SNe).

%predictions from Nadia's paper
Recently, a number of peculiar nuclear transients were discovered by various surveys (e.g. \citealt{2016NatAs...1E...2L,2017ApJ...844...46B,2017NatAs...1..865K,2017MNRAS.465L.114W}). Some of these transients were not discovered by AlertPipe, even though the sources had been detected by {\it Gaia}. Part of the goal of this paper is to investigate the reasons for this. The predicted number of detected SNe is around 1300 per year assuming a 19 magnitude minimum threshold for the brightness of the transient. About 15 per cent of these are predicted to occur in the host nuclei with offsets smaller than 1 arcsec. Moreover, $20\pm1$\footnote{The value of uncertainty from \cite{2016MNRAS.455..603B} was corrected (private communication). Poisson noise was not taken into account in the simulations by \cite{2016MNRAS.455..603B}.} TDEs should be discovered every year (\citealt{2016MNRAS.455..603B}).
From mid-2016 to mid-2017 - when a stable version of AlertPipe was operating - GSA detected and published about 50 events preliminarily classified as nuclear transients\footnote{\tt http://gsaweb.ast.cam.ac.uk/alerts} (i.e. transients - likely SNe close to the host centre or AGN variability - observed within 0.5 arcsec from their host centre if the host is recognised using external catalogues) which is roughly less than 25 per cent of the expected number of supernovae and TDEs. We note that the predictions of  \citet{2016MNRAS.455..603B} do not include events due to AGN variability, which are a significant contributor to the published {\it Gaia} Alerts, hence the missing fraction of nuclear transients is probably significantly larger than 75 per cent.

%this paper
In this study we performed a large-scale and systematic search for transient events in the nuclei of galaxies detectable by {\it Gaia} between mid-2016 and mid-2017. We started with objects classified as "galaxy" by the Sloan Digital Sky Survey Data Release 12 (SDSS DR12, \citealt{2015ApJS..219...12A}). We used a different method to search for transients than the AlertPipe daily search (\citealt{2012IAUS..285..425W,2013RSPTA.37120239H}, Hodgkin et al. in prep.). This paper is organized as follows. In Section 2 we introduce our data sample and present our transient-selection method. Next, we describe the newly found candidate nuclear transients in Section 3. In Section 4, we discuss our results and implications for {\it Gaia} Science Alerts. We conclude in Section 5. Throughout this paper we assume a flat $\Lambda$-Cold Dark-Matter ($\Lambda$CDM) concordance cosmological model of the Universe with parameters $\Omega_\Lambda = 0.7$, $\Omega_\mathrm{M} = 0.3$ and $H_0 = 70\mathrm{~km~s^{-1}~Mpc^{-1}}$, $h = 0.70$.

\section{Method}
\label{sec:method}

%Gaia data description 
The Astrometric Field (AF) instrument in the focal plane of {\it Gaia} contains 7 rows with 8 or 9 CCDs each. During a scan the position and brightness of an object is measured in each CCD that it crosses due to the motion of the satellite. This means that each time a source passes the {\it Gaia} focal plane 9(8) data points (af1-af9), separated in time by about 4.4 seconds (e.g. \citealt{2017A&A...599A..32V}) are  obtained. This collective number of 9 or 8 data points is called a transit. All observations are taken in {\it Gaia}'s $G$--band filter --  a white-light band pass folded through the response curves of the various {\it Gaia} components (e.g. mirror and CCD responses; \citealt{2016A&A...595A...7C}). For the transient detection recipe employed here, we require that at least 8 of the CCD measurements must return a valid photometric data point. This strict filtering helps weed-out cosmic-ray induced and instrumental artefacts that can affect the photometry. We determine the median magnitude per transit from the 8--9 individual CCD measurements. All photometric data points shown in this paper are such a 8 or 9 CCD median. We build the {\it Gaia} light curve for each source by combining the measurements for the different transits for that source. 

In this study we made use of all {\it Gaia} photometric data collected from the beginning of the mission (July 2014) until the end of June 2017 (\citealt{2017A&A...599A..32V}) that are ingested into the {\it Gaia} Science Alerts Database (GSA DB). By using the GSA DB we have access to the {\it Gaia} time series and individual measurements from scans (as opposed to data available in {\it Gaia}'s early data releases where only averaged data products are available). Individual transits might be lost because they were blacklisted by Initial Data Treatment as bad data or missed due to data delivery disruptions. The data obtained between mid-2014 and mid-2017 were used to compute light curve statistics such as median, skewness etc., although we are searching for transients which occurred between mid-2016 and mid-2017. The primary sample of candidate galaxies detected by {\it Gaia} was obtained by cross-matching any galaxy-like object from the SDSS DR12 (covering roughly one third of the sky) with the GSA DB.
All extended objects with photometric classification flag ''galaxy'' (based on the object morphology; \citealt{2002AJ....123..485S}) and that are brighter than magnitude 20 in the SDSS $r$-band were used.
We use the SDSS $r$--band model magnitude given in the AB system. The model magnitude is the better of two fits i.e.~a de Vaucouleurs and an exponential model fit to the SDSS light profile (\citealt{2002AJ....123..485S}). The {\it Gaia} $G$-- band magnitudes in the GSA DB are derived from a preliminary calibration of the photometry and are on the VEGAMAG system (\citealt{2016A&A...595A...7C}). The uncertainty of each data point in the light curve (each transit) is calculated using the median absolute deviation from af1-af9 measurements. The scatter for the individual light curves is also calculated using the median absolute deviation (from all transits in the light curve). 
%http://www.sdss.org/dr12/algorithms/magnitudes/#mag_model
%http://www.sdss.org/dr12/algorithms/classify/

\begin{figure}
\includegraphics[width=\columnwidth]{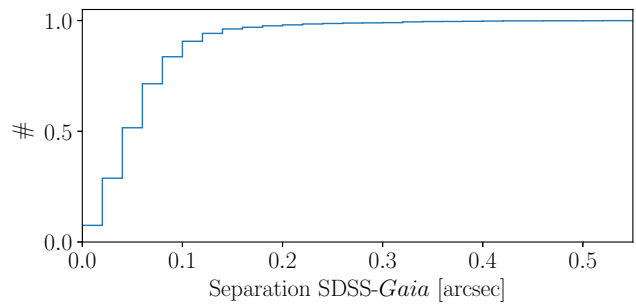}
\caption{Cumulative histogram of the separations between the SDSS coordinates of the subsample of spectroscopically confirmed galaxies and the {\it Gaia} coordinates of associated sources. The distance is typically lower than 0.5 arcsec which is our current requirement for inclusion.}
\label{fig:sdsssep}
\end{figure}

First, we attempted to identify a {\it Gaia} source with the subsample of the SDSS spectroscopically confirmed galaxies by cross-matching on source position. 
In Fig. \ref{fig:sdsssep} the histogram of the separations between SDSS and associated {\it Gaia} sources is presented. The distance is usually lower than 0.5 arcsec. Hence, we assumed that an SDSS source falling in a circle with a radius of 0.5 arcsec around a GSA DB source implies that the two sources are the same (taking into account the inaccuracy in detecting the galaxy centres by SDSS and {\it Gaia}'s astrometric uncertainty estimated at 100 mas obtained from the Initial Data Treatment, \citealt{2016A&A...595A...3F}). From the 18.5 million SDSS objects fewer than 4 million have one or more counterpart source(s) in the GSA DB. 

One should note that the SDSS sample is strongly contaminated by unresolved binaries and spurious source detections caused by diffraction spikes from bright stars. However, the number of real galaxies detected by {\it Gaia} is reduced mainly due to the on-board data detection algorithm (\citealt{2015A&A...576A..74D}). The detectability of galaxies is a function of their brightness, size, and compactness (see \citealt{2014sf2a.conf..421D,2015A&A...576A..74D,2016MNRAS.455..603B}).

\begin{figure}
\includegraphics[width=\columnwidth]{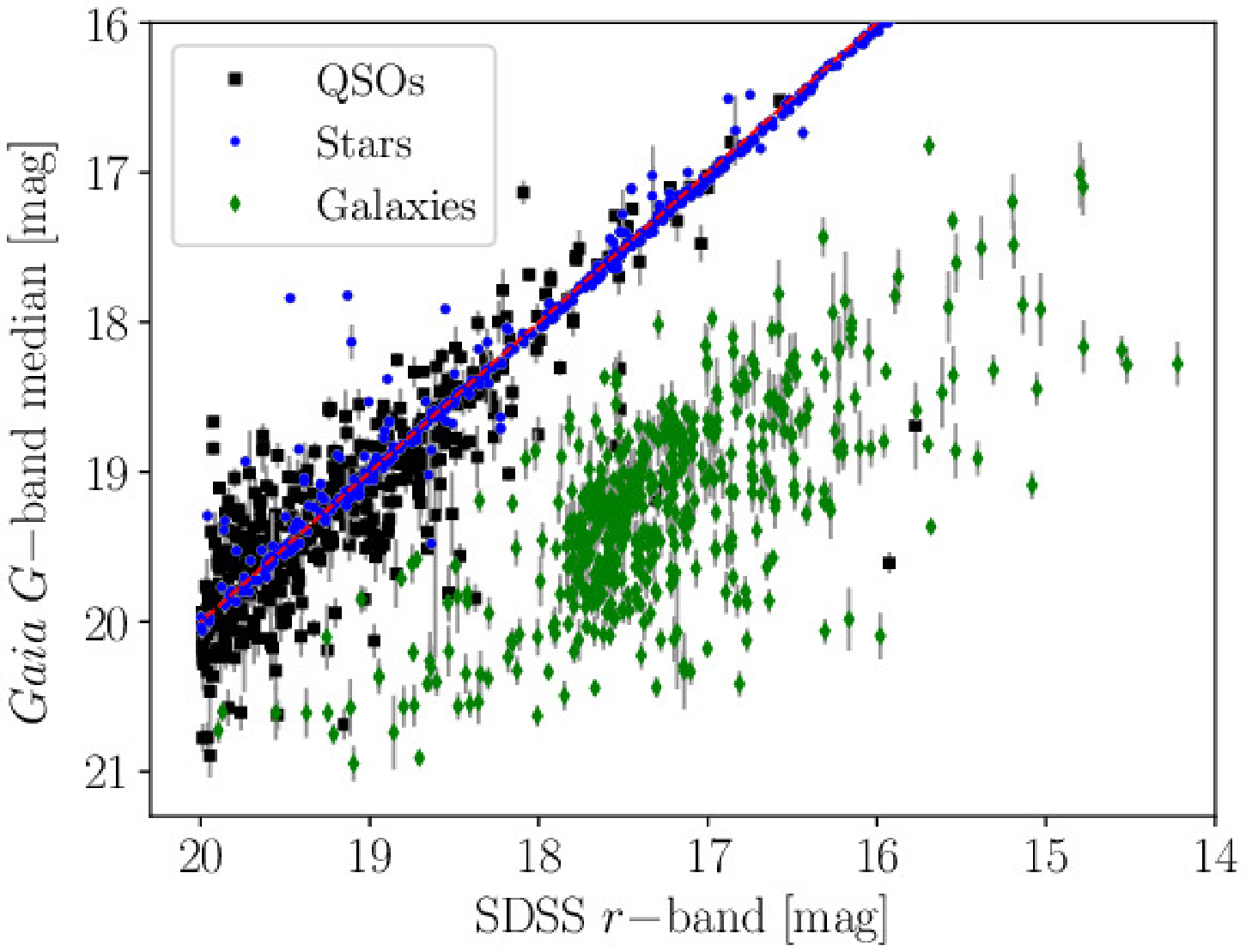}
\includegraphics[width=\columnwidth]{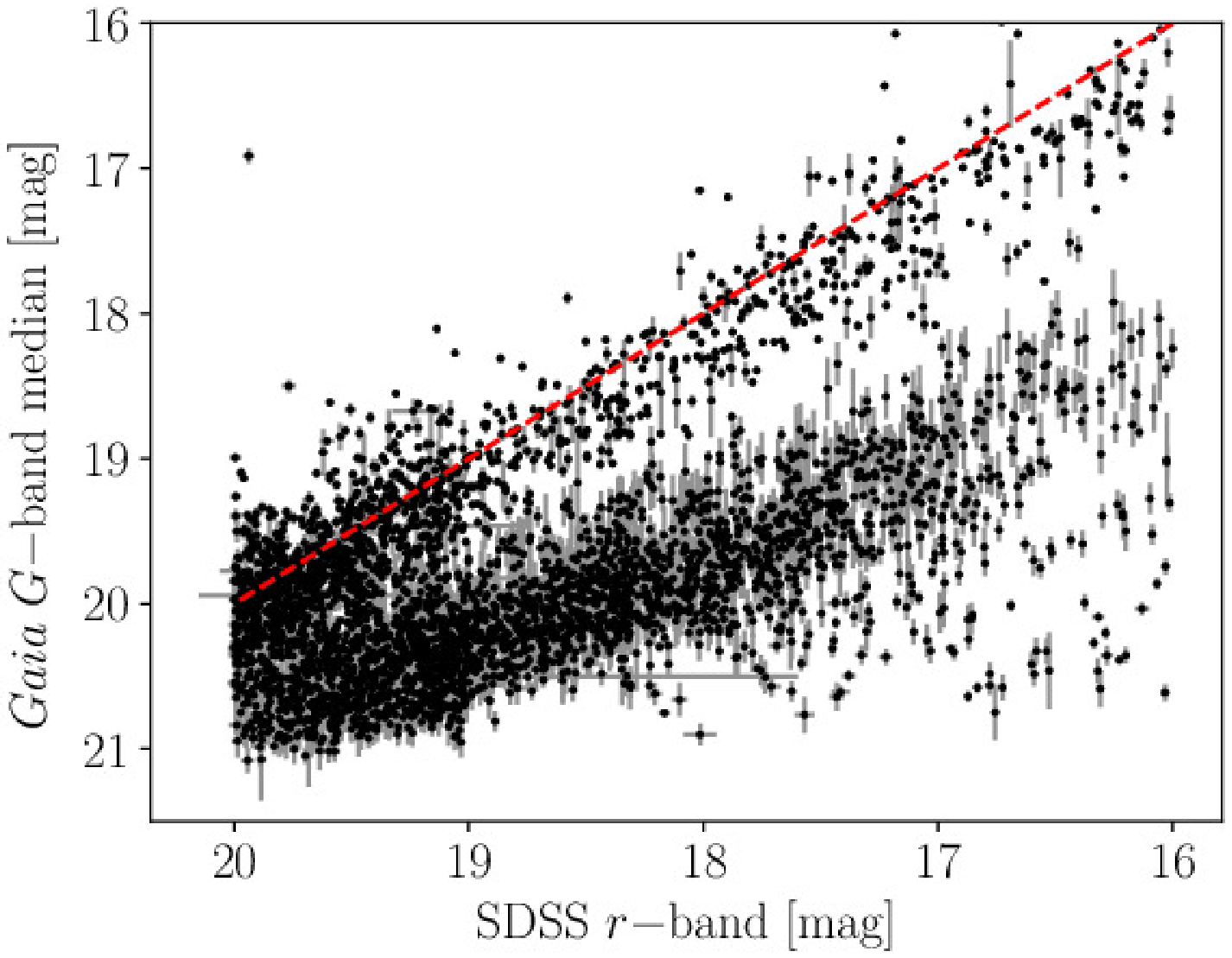}
\caption{Comparison of the magnitude of the same source detected by {\it Gaia} ($G$--band median) and SDSS ($r$--band model magnitude). {\it Top:} A sample of spectroscopically confirmed objects from SDSS - quasars (black squares), galaxies (green diamonds), and stars (blue dots) cross-matched to the sources in GSA DB. The {\it Gaia} detections of the extended SDSS sources (galaxies) return typically fainter magnitudes in comparison to the SDSS brightness. The {\it Gaia} light curves have at least 10 data points. The red dashed line indicates 1:1. {\it Bottom:} A subset of all detected sources (from the SDSS subsample of galaxy-like objects) before our data filtering has been applied (we randomly chose 4000 out of 4.5 million objects for the figure). The sources on the upper branch in the diagram are probably point sources such as QSOs and binaries whereas real extended galaxies fall on the lower branch. The red dashed line indicates 1:1.}
\label{fig:magsdss}
\end{figure}
\subsection{Filtering sources and light curves}
\label{sec:filters}
In Figure \ref{fig:magsdss} (top panel) we compare the brightness in the $r$--band filter of the randomly chosen SDSS sources spectroscopically classified as quasars, galaxies, and stars to the median brightness of the matched {\it Gaia} objects in the $G$--band. As can be expected, the {\it Gaia} sources are typically fainter than the SDSS detections of the same extended objects. The main reason for this is that {\it Gaia} sums the light of these extended objects over a much smaller angular region than SDSS. {\it Gaia}'s on-board flux measurement algorithm assumes that the source has a point-source-like profile and using the estimated brightness sets a window with a fixed size. For sources with $20>G>16$ the window has a size of 12$\times$12 CCD pixels (\citealt{2015A&A...576A..74D}), where the pixels in the along scan direction have a size of 59 mas and the pixels across the scan direction have a size 3 times larger (i.e.~177 mas). Hence, the window size for extended objects is typically smaller than the size of the source. Another, secondary, reason for the different magnitudes lies in the differences in bandpass between the SDSS $r$-- and the {\it Gaia} $G$--band. In Figure \ref{fig:magsdss} (bottom panel) the brightness of the objects (extended according to SDSS and photometrically classified as galaxies) in the $r$--band filter is compared to the median brightness of the matched {\it Gaia} objects in the $G$--band. 

The sample obtained from the cross-match was filtered using various selection criteria.
\begin {enumerate}
\item Criteria related to the SDSS sample quality: 
\begin {enumerate}
\item  We employed a colour-colour selection on the SDSS data. We require the source colours to be $-0.5<g-r<2.0$ and  $-0.5<r-i<0.8$ to remove binary systems, bright stars, and diffraction spikes from bright stars, that are all misclassified as a galaxy (\citealt{1999ApJS..123..377N,2001AJ....122.1861S}).
\item We removed all objects spectroscopically classified as stars.
\end {enumerate}
\item Criteria to the {\it Gaia} data:
\begin {enumerate}
\item The standard deviation within the measurements from the 9(8) CCD detectors of all data points is less than 0.25 mag (because we assume no significant change in the light curves within the 45 seconds that the source transits over the focal plane). We only remove bad data points and do not reject the source unless as a result it has less than 10 remaining data points.
\item The {\it Gaia} light curve must have at least 10 good data points (i.e. 10 transits) up to the maximum brightness of the light curve to have a sufficiently large sample of data points to perform our statistical studies (as a reminder, 1 data point consists of the median of 8 to 9 valid CCD measurements). This requirement is the same as GSA uses in their {\it OldSource} transient detection algorithm.
\item We require that the peak of the {\it Gaia} light curve falls between 2016 July 1 and 2017 June 30 to be able to compare the results with the stable version of AlertPipe.
\item The amplitude of the rise from the baseline of the light curve exceeds 0.3 mag (measured from the median value determined using all the data points in the light curve to the maximum). 
\item We require that there be no additional {\it Gaia} source detections within a range of 0.3-3 arcsec for the majority of detections in the light curve of the candidate transient. Using this criterion we aim to remove binary systems that are unresolved by SDSS but that are resolved by {\it Gaia}. 
\item We require that there be no {\it Gaia} DR1 star brighter than $G=14$ mag within a circle of a radius 25 arcsec (\citealt{2016A&A...595A...1G,2016A&A...595A...2G}), to remove any remaining spurious detections in the GSA DB due to diffraction spikes of this bright source.
\item Occasionally {\it Gaia} flux measurements are affected by unexpected events (for example solar flares, spacecraft or processing artefacts). This may cause an erroneous excess in flux not corrected for by the Initial Data Treatment (although this is corrected later during the data processing for main {\it Gaia} Data Releases, it is not corrected for AlertPipe and affects the GSA DB). We analysed the number of candidate transients found by our search in time intervals of 0.5 days. We found that during some periods the number of candidate transients with high amplitude outbursts is unusually large. We plotted the number of candidate transients in 0.5 day bins weighted by the amplitude squared as a function of time. We calculated the mean and standard deviation and disregarded the epochs $>$ 3 sigma from the mean as the exact cause of these anomalous flux measurements for some sources is not known at present. In total consecutive 4 periods of half a day of data were discarded.
\end {enumerate}
\end {enumerate}

The impact of each selection criterion applied during filtering on the sample size is summarised in Tab. \ref{tab:cri}.
\begin{table}
\centering
\caption{A summary of the impact of each selection criterion applied during filtering on the sample size. The initial sample size is $3.96\times10^6$. The table provides the number of SDSS sources and associated {\it Gaia} objects - these numbers differ as multiple {\it Gaia} objects might be associated with one SDSS object.}
\begin{tabular}{l l l}
\hline
Criterion & \# of SDSS objects & \# of {\it Gaia} objects \\
\hline
(ia) & $3.40\times10^6$ & $3.78\times10^6$\\
(ib) & $3.39\times10^6$ & $3.77\times10^6$\\
(iib)+(iic) & $0.53\times10^6$ & $0.55\times10^6$  \\
(iid) & $61.6\times10^3$ & $64.2\times10^3$ \\
(iie) & $55.2\times10^3$ & $57.6\times10^3$ \\
(iif) & $32.1\times10^3$ & $32.3\times10^3$ \\
(iig) & $32.0\times10^3$ & $32.2\times10^3$ \\
\hline
\end{tabular}
\label{tab:cri}
\end{table}

\subsection{Simulation of light curves}
In our search for transients we applied a novel detection algorithm.
We calculate the von Neumann statistic -- the ratio of the successive mean square difference to the variance: 
\begin{equation}
\eta=\frac{\frac{1}{n-1}\sum_{i=1}^{n-1}(x_{i+1}-x_i)^2}{s^2};
\end{equation} 
(\citealt{vonneumann1941}) as well as the skewness of the light curve: 
\begin{equation}
\gamma=\frac{\frac{1}{n}\sum_{i=1}^{n-1}(x_i-\bar{x})^3}{s^3}
\end{equation}
(where $x$ are the flux measurements during transits, $s$ -- the variance of the light curve, $n$ -- the number of transits in the light curve).
These two statistics were previously used in searches for e.g.~microlensing events (\citealt{2014ApJ...781...35P, 2016MNRAS.458.3012W}), eclipsing binaries (\citealt{2015MNRAS.447L..31R}), and fast {\it Gaia} transients (\citealt{2018MNRAS.473.3854W}). 

\begin{figure}
\includegraphics[width=\columnwidth]{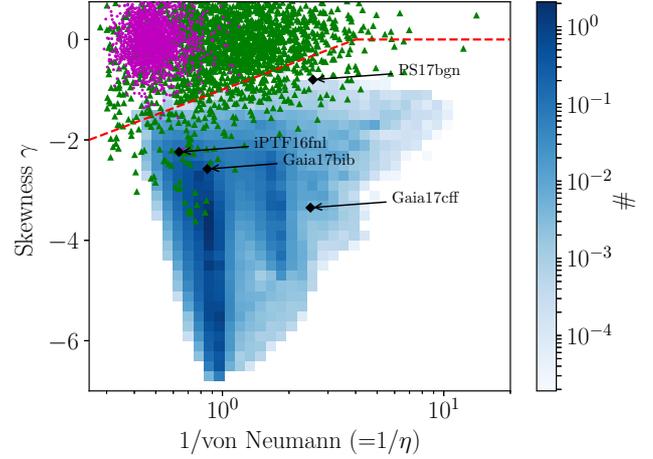}
\caption{The skewness vs. the reciprocal of von Neumann statistic for simulated light curves. The blue squares show the parameter space for the simulated light curves with ingested transients (supernovae Ia, Ibc, IIL, and IIP and peculiar transients such as ASASSN-14li, ASASSN-15lh, iPTF16fnl). However, different types of transients do not populate different regions on this parameter space. Magenta points are the simulated constant flux light curves with noise. Green triangles represent the quasar light curves simulated with the damped random walk model. Black diamonds indicate known nuclear transients and their position on the skewness -- the reciprocal of von Neumann parameter plane. The red dashed lines indicate the applied cuts $(\gamma<0~\mathrm{and}~\gamma<\log(1/\eta)/\log(4)-1)$. The cut ensures that 99 per cent from the transients will be detected, while only $\sim$10 per cent false positives (quasar light curves with stochastic variability) will pass the criteria. Number of true and false positives can be used as a parameter in the detector to reduce the eye-balling but at the expense of losing real events.}
\label{fig:vns_sim}
\end{figure}

Using the predicted data sampling from the Nominal Scanning Law we simulated {\it Gaia} light curves of galaxy centres located in various parts of the sky. The noise model for light curves without transients comes from the observed noise in real galaxies observed as in Fig. \ref{fig:magsdss}. Template light curves for supernova Ia, Ibc, IIL, and IIP (\citealt{2002PASP..114..803N}; {\tt https://c3.lbl.gov/nugent/nugent\_templates.html}) and real light curves for TDE/SLSN candidates (ASASSN-14li \citealt{2016MNRAS.455.2918H}, ASASSN-15lh \citealt{2016Sci...351..257D,2016NatAs...1E...2L}, iPTF16fnl \citealt{2017ApJ...844...46B}) were used to simulate the expected light curves of transients occurring on top of galaxies. Using uniform distributions we randomised galaxy core brightness (15-21 mag), transient brightness (15-21 mag), transient redshift used only for stretching the observed light curves (0-0.5), and time of transient maximum (mid-2016-mid-2017). We also simulated constant flux light curves with noise and stochastic variability observed in quasars, described by the damped random walk (DRW) model (\citealt{2010ApJ...708..927K,2010ApJ...721.1014M}. For the DRW model we used the normal distributions from  \cite{2010ApJ...721.1014M}  ($\tau=\mathrm{normal}(\mu=2.4,\sigma=0.2),\mathrm{SF_{\infty}}=\mathrm{normal}(\mu=-0.51,\sigma=0.02$)). For each obtained light curve from our Monte Carlo simulation we calculated the skewness and von Neumann statistics using identical constraints for light curves as described in Subsection \ref{sec:filters} (e.g. the light curve must have at least 10 data points up to the maximum, the amplitude of the light curve exceeds 0.3 mag). Investigating a plot of the skewness vs. the reciprocal of von Neumann statistic, we noticed that all simulated light curves with transients occupied one part of parameter space (see Fig. \ref{fig:vns_sim}) and based on this we chose limits for skewness and von Neumann statistics where one is most likely to find a nuclear transients candidate. As a possible alternative, we also investigated the use of Gaussian Mixture Models to separate various types of simulated light curves (\citealt{2014sdmm.book.....I}). However, only the simulations of constant flux light curves with noise were well separated whereas the overlapping samples of transients and quasar light curves stayed undivided from each other and this approach was rejected. The exact ratio of transients to quasars will depend on the relative sizes of these populations that is not included in the simulations. 

\subsection{Statistical analysis of {\it Gaia} light curves}
\label{sec:stats}
\begin{figure}
\includegraphics[width=\columnwidth]{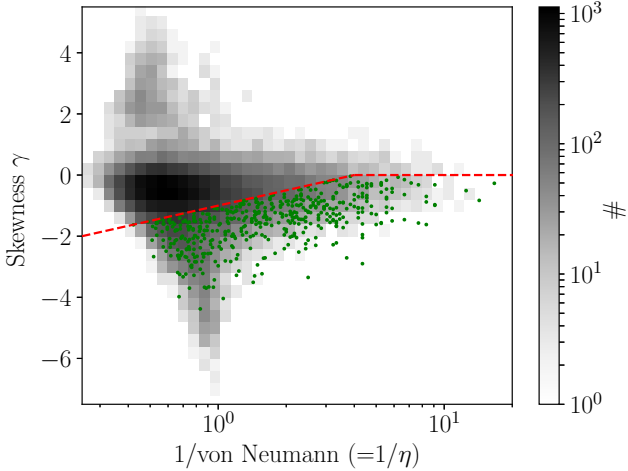}   
\caption{The skewness vs. the reciprocal of von Neumann statistic of the light curve for all (filtered) {\it Gaia} detections of SDSS DR12 galaxies. The grey squares in the background show the distribution from the whole sample after applying the filters described in Subsection \ref{sec:filters}; the sample contains $\sim$32k objects. The green points indicate the selected subsample with transient candidates. We only study light curves with $\gamma<0$ as skewness greater than zero means that the light curve is declining or a dip occurs in the light curve.} After applying an additional constraint on the skewness -- von Neumann plane (see Subsections \ref{sec:stats} and \ref{sec:addch}) the sample shrunk to $\sim$7k objects (which was limited to $\sim$6k by using complete light curves). Due to eyeballing $\sim$480 ($\sim$8 per cent) sources remained as transient candidates (green points). The red dashed lines indicate the applied cuts $(\gamma<0~\mathrm{and}~\gamma<\log(1/\eta)/\log(4)-1)$.
\label{fig:vns}
\end{figure}

This search was performed on historic data rather than using the daily ingest and processing  as AlertPipe does.
For all {\it Gaia} light curves we computed the von Neumann statistic and skewness using data points up-to and including the light curve maximum. Therefore we only use the part of the data, which will at most be available if such a statistic is implemented in a near-real time transient search, for instance in a future version of AlertPipe. Besides, the uneven {\it Gaia} sampling can cause part of the light curve to be sampled relatively frequently and if this happens to correspond to the outbursting phase, then a median over all data would render the outburst undetectable. In Figure \ref{fig:vns} we present the measurements for all the sources (filtered using the criteria from Subsection \ref{sec:filters}). The whole sample after applying these filters contains $\sim$32k objects. 

Our simulations informed which part of the skewness -- the reciprocal of von Neumann parameter space is most relevant for our search of transients. We decide to exclude sources in the skewness -- the reciprocal of von Neumann plane if they falls above the line drawn in Figure~\ref{fig:vns_sim}. After applying that constraint on the skewness -- the reciprocal of von Neumann plane the sample shrunk to 7k objects.

\subsection{Additional checks on the transient candidates}
\label{sec:addch}

\begin{figure}
\includegraphics[width=\columnwidth]{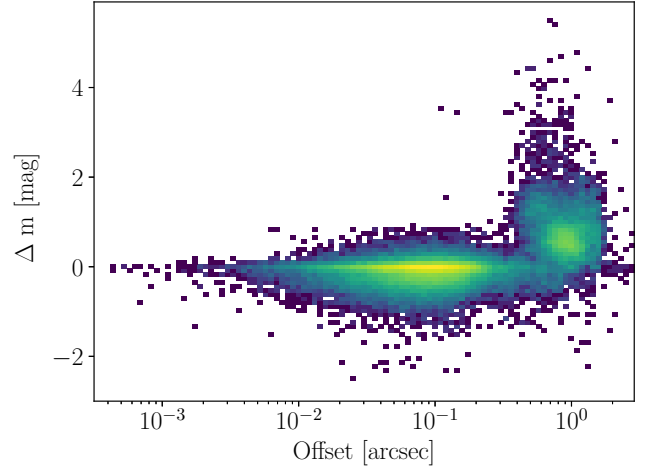}
\caption{The plot shows the difference in magnitude of sources whose position falls at least once within a cone search with a radius of 1.5 arcsec around the median position of the transient candidate, and the median magnitude of the transient candidate. Two clusters of points are apparent - one with offsets between 0.5--2 arcsec with differing magnitudes and one with on offset centred at 0.1 arcsec with magnitudes consistent with the median magnitude of the candidate transient. The former group of measurements probably belongs to real objects unrelated to the transient candidate. The latter group of measurements are probably erroneously assigned to another source ID during the Initial Data Treatment (\citealt{2016A&A...595A...3F}), whereas they are most likely associated with the candidate transient.}
\label{fig:neigh}
\end{figure}

It is known that the {\it Gaia} initial data processing on occasion assigns a different source identification number (source ID) to the same source if the position is apparently shifted with respect to the historic position (\citealt{2016A&A...595A...3F}). This would then cause the light curve information of that source to be split over more than one source ID making our transient detector less sensitive as less data points per astrophysical source are available than if only one source ID per astrophysical source is used. To investigate if this happens, we plot in Figure \ref{fig:neigh} on the ordinate the difference in the magnitude of sources whose source position falls at least once within a cone search around the median position of the candidate transient with a radius of 1.5\arcsec\, and the median magnitude of the  candidate transient. On the abscissa we plot the separation between the median position of our candidate transient and the median position of such a source.

Two clusters of points are apparent. First, there is a group of transits with offsets between 0.5--2\arcsec\ with magnitudes lower than the median magnitude of the candidate transients. We can filter such occurrences out by checking for source entries that have a similar coordinate in {\it Gaia}'s detector plane. This leads to a filter on the difference between the coordinates in the across--scan direction of the two sources. The main group of sources is centred at an offset of 0.1\arcsec\, and at a magnitude difference with respect to that of the median of the candidate transient consistent with zero. These characteristics are consistent with the situation where the Initial Data Treatment erroneously assigned more than one source ID to a single astrophysical object. For instance, the $\sim$0.1\arcsec\, offset is similar to the offset of the mean distance of 0.06$\pm0.05$\arcsec\ between SDSS galaxies and {\it Gaia} DR1 counterparts and reflects the astrometric accuracy of the {\it Gaia} data one day after data taking. Similarly, along the ordinate, the standard deviation in the magnitude difference for the different source IDs of 0.4 magnitude is consistent with the standard deviation of the magnitude measurements of the candidate transients before 2016 (which is 0.36 mag). The similarities in these distributions support the idea that these measurements are drawn from the same distribution of measurements as that of the sources that led to the identification of the candidate transients.

Therefore, we combined the magnitude measurements of the various source IDs that belong to the same astrophysical source. Hereafter, for each astrophysical source, we have more entries in the light curve than before, and hence, now we have an improved handle on the source location in the skewness -- the reciprocal of von Neumann plane. We recalculated the von Neumann and skewness values for these new, more complete, light curves and we reapplied our selection criteria in the skewness -- the reciprocal of von Neumann plane. This way we deselect about 1000 sources that now fall in the group of variable sources instead of that of the transient sources. 

As a last step we visually inspected the {\it Gaia} light curves and the SDSS finding charts of the candidate transients.
The whole sample after applying the filters described in Subsection \ref{sec:filters} contains 32236 objects. After applying all additional constraints the sample shrunk to 6091 objects from which due to eyeballing 804 ($\sim$15 per cent) sources remained as candidate transients. Usually false positives were binary systems unresolved by SDSS, objects close to bright stars or QSOs with stochastic variability. However, for the half of the sample (423 candidates) the transients are a single transit events that may be caused by parasitic sources (see \citealt{2018MNRAS.473.3854W}). These candidates are most likely due to an effect that is caused by the fact that the field of view of both {\it Gaia's} telescopes is projected on the same focal plane. When one of the two telescopes points at regions of the sky that are densely populated with stars, such as the Galactic plane, there is a risk that the regions centred on the sources on the detector that {\it Gaia}'s on-board detection algorithm sets start to overlap. If this happens, (part of) the flux of one source (always the fainter of the two) is added to that of the other artificially brightening it. Hence if the transient event occurs and at the same time the second field of view is observing the area closed to Galactic plane ($|b|<30^\circ$) we exclude the candidate from the list (we rejected 322 sources from 423 candidates). After this last selection we are left with 482 sources in total. 

\section{Results}
\subsection{New candidate nuclear transients}

\begin{figure*}
\begin{tabular}{cc}
\includegraphics[width=80mm]{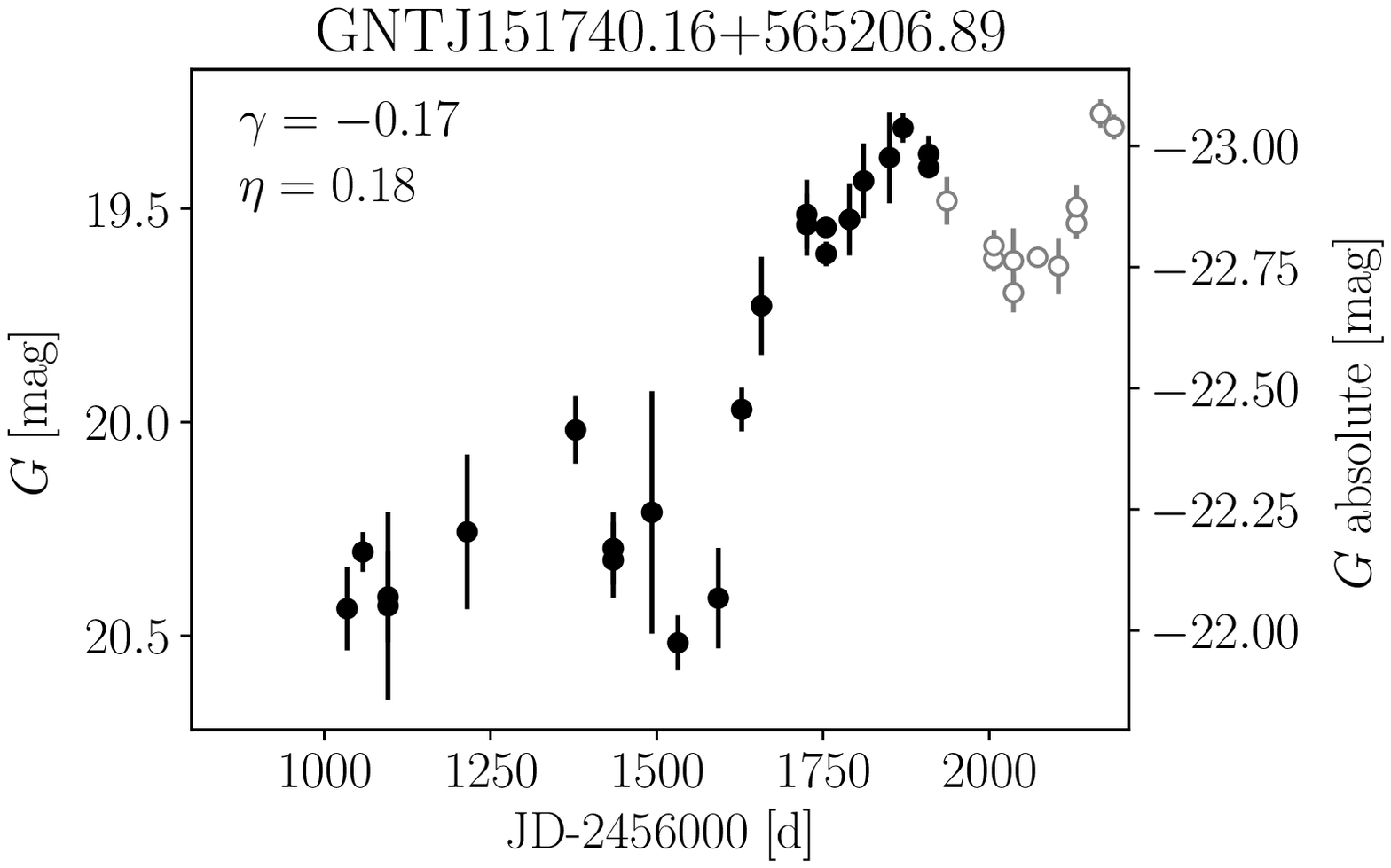} & \includegraphics[width=80mm]{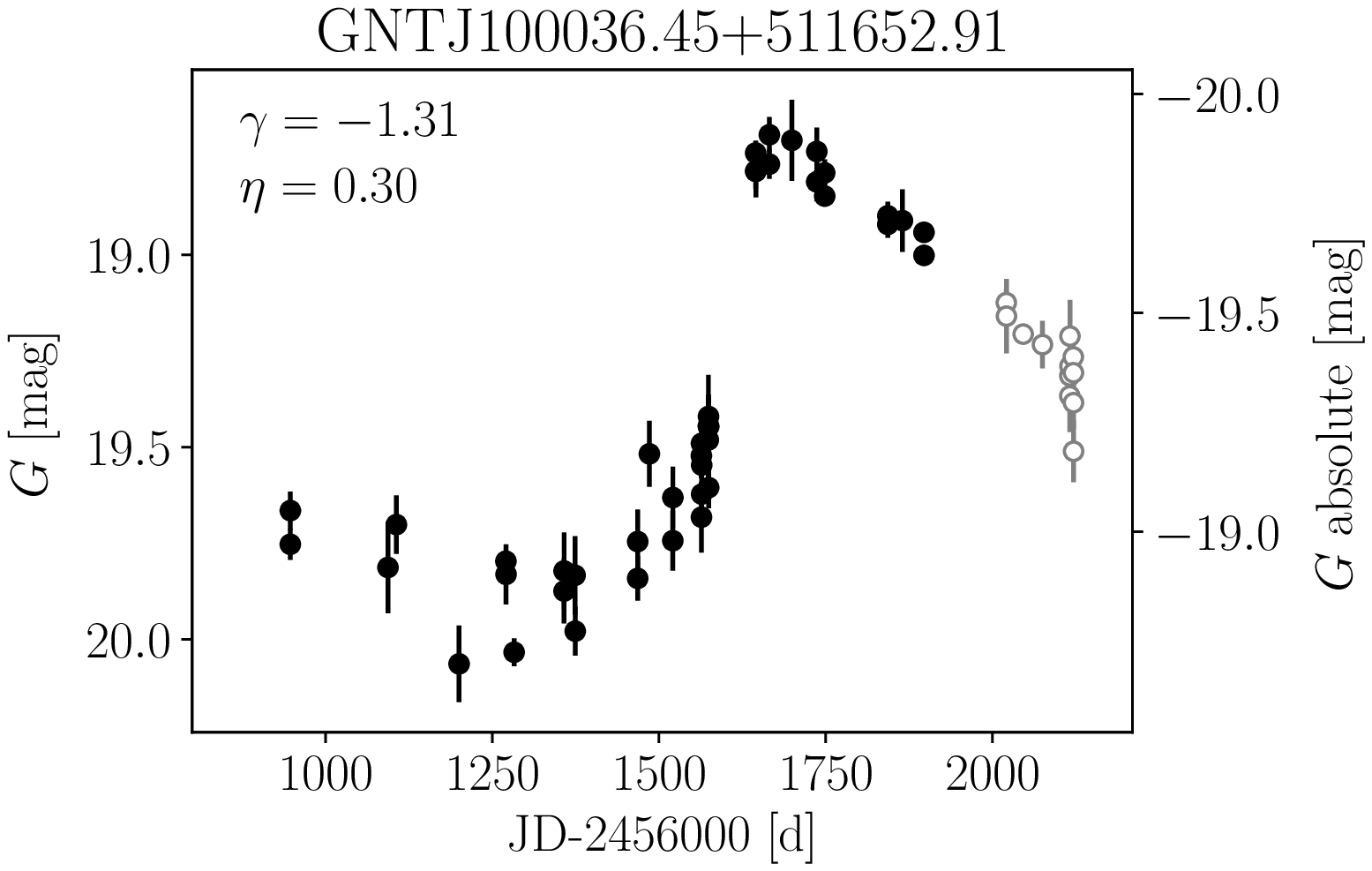} \\
\includegraphics[width=80mm]{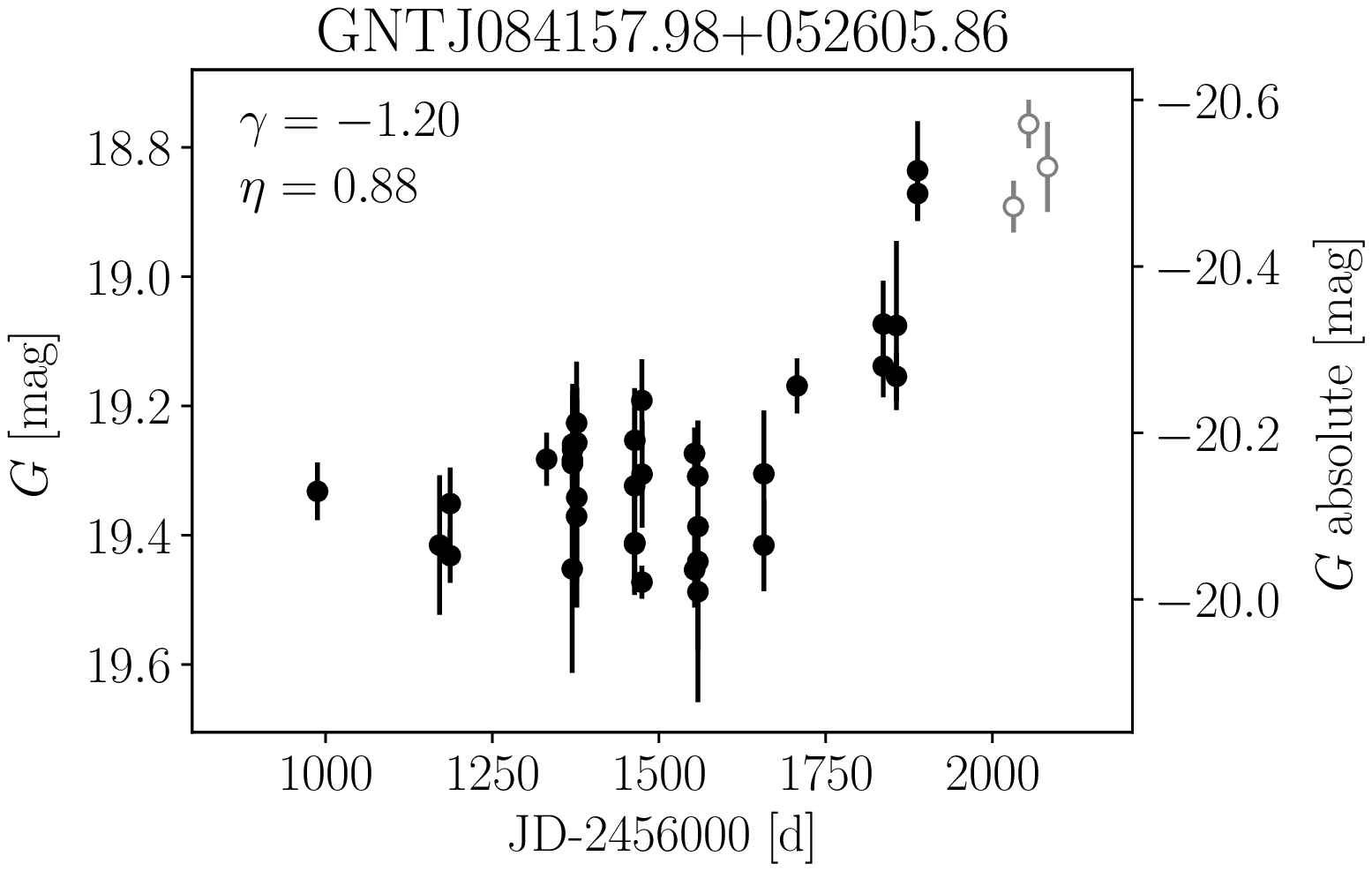} & \includegraphics[width=80mm]{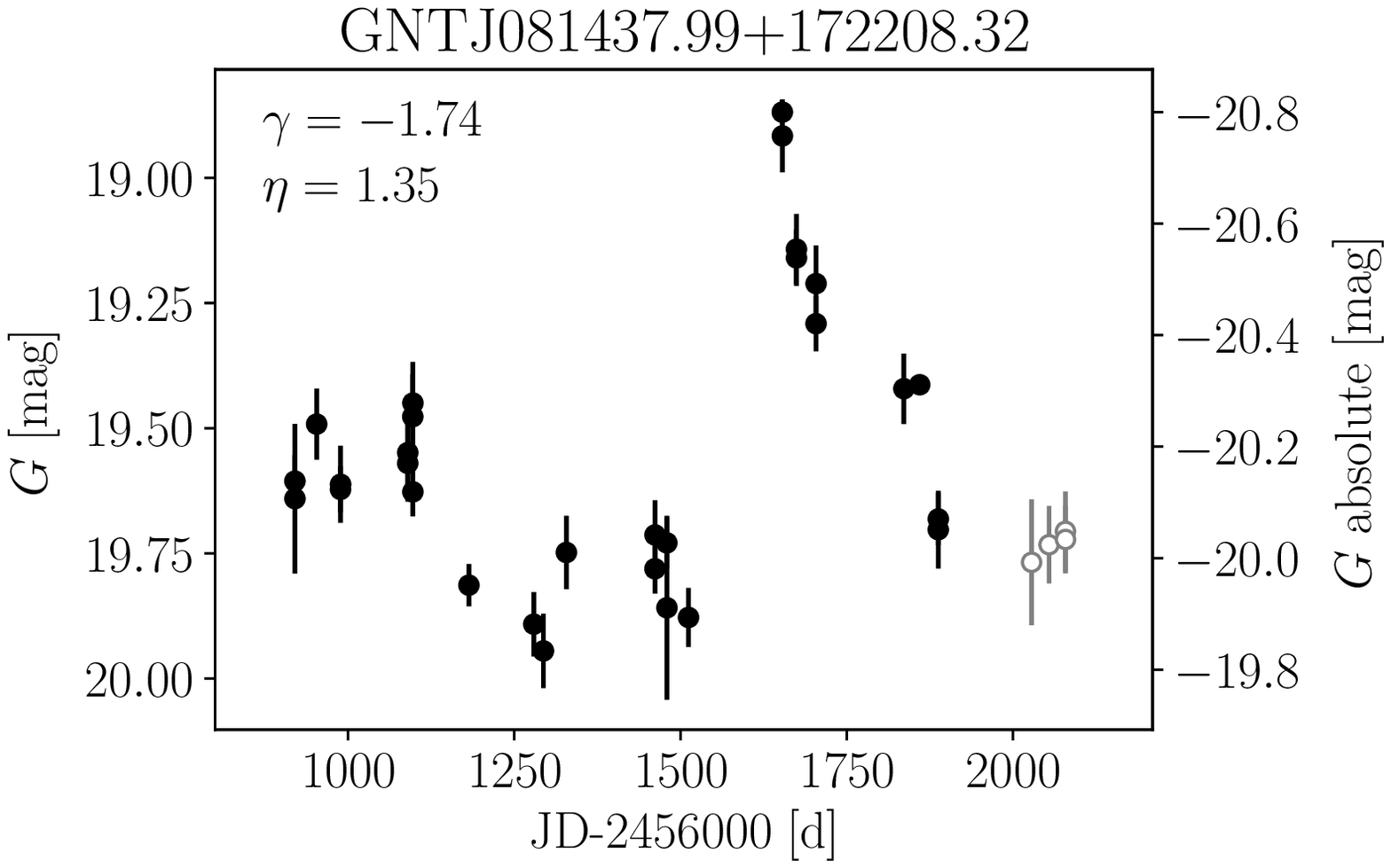} \\
\includegraphics[width=80mm]{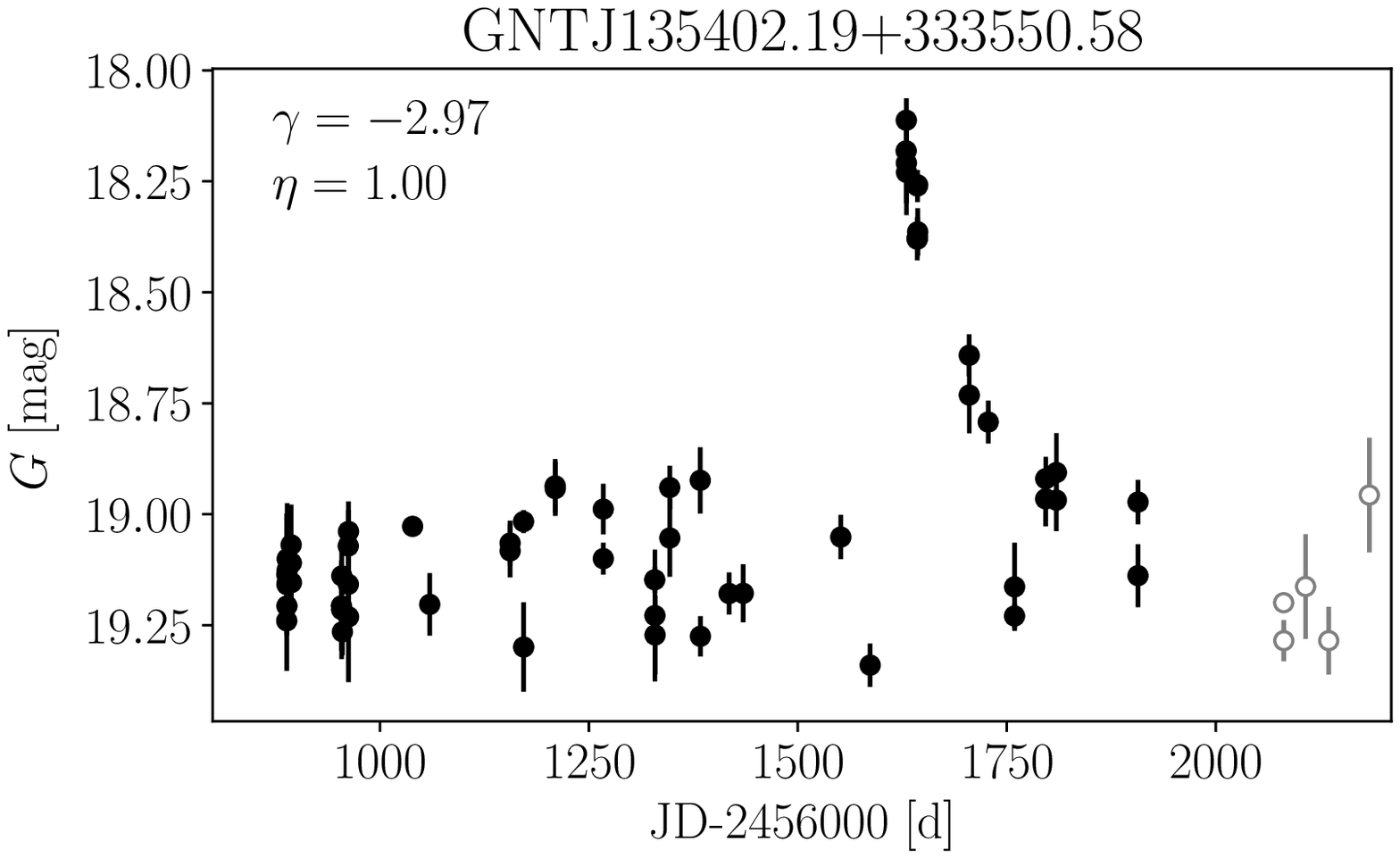} & \includegraphics[width=80mm]{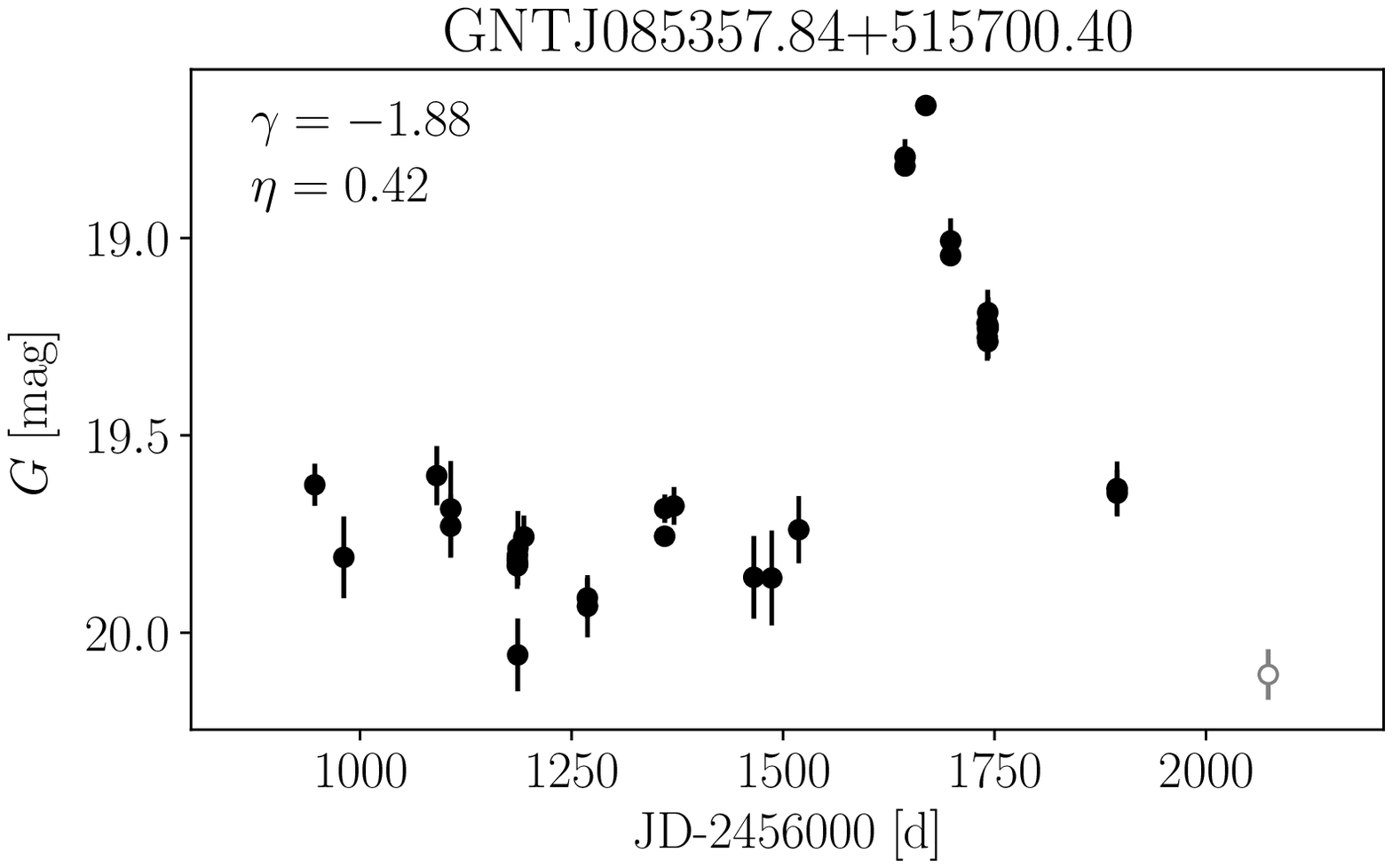} \\
\end{tabular}
\caption{Examples of light curves for candidate {\it Gaia} nuclear transients. The skewness $\gamma$ and von Neumann $\eta$ parameters are provided. The grey open circles denote {\it Gaia} data collected after the period that we employed for our transient search. The absolute magnitudes are provided if spectroscopic redshift is available. Note: The plots and light curves in ascii format for all objects are available from the online journal.} 
\label{fig:lcgnt}
\end{figure*}

\begin{figure}
\includegraphics[width=\columnwidth]{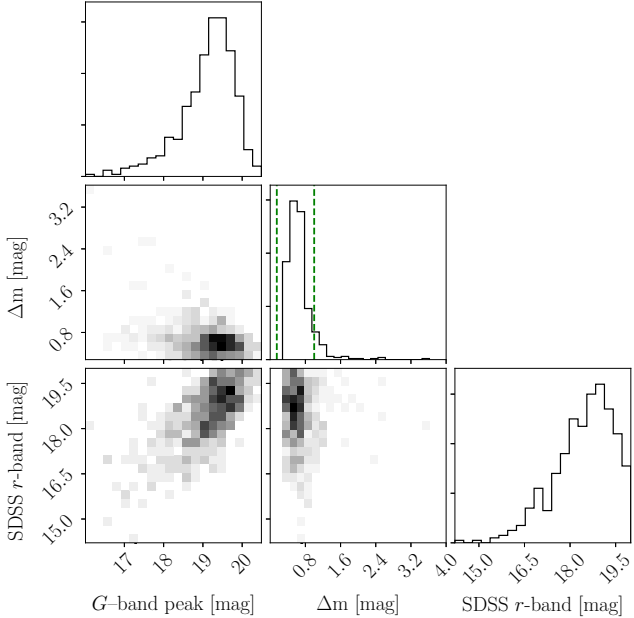}
\caption{A corner plot of the amplitude of the rise of the transients in the {\it Gaia} light curve ($\Delta$m) vs.~the {\it Gaia} $G$--band magnitude at light curve maximum vs.~the SDSS $r$--band magnitude. The maximum brightness of the flares spans between 16 and 20.5 mag in the {\it Gaia} $G$--band. About 160 transients are brighter than 19$^{th}$ mag and $\sim$290 that have a peak magnitude between 19 and 20 mag. Only 9 per cent of transient candidates have an amplitude of the rise greater than 1.0 mag (which is a requirement on one of the two ways in which the {\it OldSource} detector can indicate a transient event in the daily GSA system, both thresholds for the {\it OldSource} detector are indicated with the dashed green lines on the $\Delta$m histogram). The amplitude of the transient event in the {\it Gaia} light curve in magnitude spans over up to 3.62 mag and there is no dependence on the source brightness in the SDSS $r$-band. Note that the lack of bright transients in bright SDSS sources ($r<16$ mag) is consistent with being due to low number statistic in these magnitude bins.}
\label{fig:histmmaxvsamp}
\end{figure}

The final sample of selected candidate nuclear transients consists of about $\sim$480 sources.
In Table \ref{tab:gnt} we provide the list of {\it Gaia} Nuclear Transient (GNT) candidates with {\it Gaia} coordinates, $G$--band median brightness and the corresponding SDSS sources with the spectroscopic class and redshift (if available). Example light curves are presented in Fig. \ref{fig:lcgnt}. The light curves contain all available data points, however the points after the peak are only plotted for informative purposes and to show how the transients are evolving. These data were not used to obtain the light curve statistics and sometimes our peak is only a local maximum (up to mid-2017) as the transients might rise further. The amplitudes of the light curves from {\it Gaia} (the difference between the light curve maximum and the median brightness in {\it Gaia}, i.e. $\Delta$m) span over 4 mag. The median brightness was determined using all the data up to the maximum. The
transient sample is not homogenous and the range of rise times is wide. Removing the data from the rise would require preparing multiple detectors sensitive for different types of transients and rise times. We noticed that the von Neumann and skewness statistics are certainly capable in finding both fast and slow rise transients.
There is no dependence of $\Delta$m on brightness in SDSS (see Fig. \ref{fig:histmmaxvsamp}). 
The detected peak brightness of the flares span between 16.01 and 20.49 mag in the {\it Gaia} $G$--band. There are about 160 sources that have a peak magnitude above 19 mag and $\sim$290 that have a peak magnitude between 19 and 20 mag (see Fig. \ref{fig:histmmaxvsamp}). 

\begin{landscape}                   
\begin{table}
\centering
\caption{The list of {\it Gaia} Nuclear Transient (GNT) candidates. The source list provides the GNT source name, the SDSS galaxy ID, the coordinates RA and Dec from {\it Gaia} in decimal degrees, the SDSS $r$--band brightness, the {\it Gaia} $G$--band median brightness, the {\it Gaia} $G$--band peak brightness, the JD at the time of the peak, the skewness parameter $\gamma$, the von Neumann $\eta$ parameter (see the text on how these are calculated), and the SDSS spectral classification with redshift (if available).
Note: This table is available in its entirety in a machine-readable form from the online journal and Centre de Donnees astronomiques de Strasbourg. A portion is shown here for guidance regarding its form and content.}
\label{tab:gnt}
\begin{tabular}{l l c c c c c c c c l}
\hline
GNT ID & SDSS galaxy ID & RA & Dec. & SDSS & {\it Gaia} & {\it Gaia}  & JD peak & $\gamma$ & $\eta$ & SDSS class \\
 & & J2000 & J2000 & $r$--band & $G$--band & peak & -2456000d & & & (redshift) \\
\hline
GNTJ000121.47$-$001140.29&SDSSJ000121.47$-$001140.32&0.33945&-0.19453&19.63$\pm$0.02&19.89$\pm$0.08&19.33$\pm$0.01&1576.45&-2.05&0.23&GALAXY(0.46152$\pm$0.00035)\\
GNTJ000426.46$+$160346.13&SDSSJ000426.46$+$160346.13&1.11023&16.06281&16.72$\pm$0.00&19.06$\pm$0.14&18.13$\pm$0.06&1767.73&-1.86&0.50&GALAXY(0.06310$\pm$0.00002)\\
GNTJ000555.83$+$101247.39&SDSSJ000555.83$+$101247.40&1.48264&10.21317&18.35$\pm$0.01&19.89$\pm$0.06&19.37$\pm$0.01&1737.00&-2.83&1.03&\\
GNTJ000725.02$+$293235.77&SDSSJ000725.03$+$293235.71&1.85426&29.54327&16.87$\pm$0.01&20.09$\pm$0.12&19.44$\pm$0.04&1738.76&-1.51&0.71&\\
GNTJ001019.55$+$025340.28&SDSSJ001019.55$+$025340.34&2.58145&2.89452&19.00$\pm$0.01&19.19$\pm$0.09&18.80$\pm$0.03&1764.73&-1.04&0.34&QSO(0.58875$\pm$0.00011)\\
GNTJ001019.96$-$061705.76&SDSSJ001019.96$-$061705.73&2.58316&-6.28493&15.74$\pm$0.00&17.81$\pm$0.13&16.97$\pm$0.03&1577.63&-1.44&1.04&\\
GNTJ001136.33$+$344830.93&SDSSJ001136.33$+$344830.90&2.90136&34.80859&19.17$\pm$0.02&19.57$\pm$0.09&19.11$\pm$0.01&1611.40&-0.84&0.77&\\
GNTJ001442.26$-$010907.08&SDSSJ001442.27$-$010907.12&3.67610&-1.15197&19.24$\pm$0.02&20.53$\pm$0.09&20.10$\pm$0.06&1714.49&-1.83&1.55&\\
GNTJ001645.72$+$105213.09&SDSSJ001645.72$+$105213.03&4.19049&10.87030&19.04$\pm$0.02&20.46$\pm$0.07&19.80$\pm$0.08&1736.42&-1.55&0.32&\\
GNTJ001917.85$-$064513.46&SDSSJ001917.85$-$064513.45&4.82437&-6.75374&17.81$\pm$0.01&19.45$\pm$0.10&19.13$\pm$0.10&1734.73&-0.91&0.78&\\
GNTJ002026.66$+$334607.55&SDSSJ002026.66$+$334607.56&5.11107&33.76876&16.71$\pm$0.01&19.30$\pm$0.16&18.60$\pm$0.05&1896.72&-2.07&0.84&GALAXY(0.12362$\pm$0.00002)\\
GNTJ002150.03$+$282847.03&SDSSJ002150.03$+$282847.03&5.45845&28.47973&16.94$\pm$0.01&19.23$\pm$0.06&18.95$\pm$0.08&1575.18&-1.48&1.75&\\
GNTJ002326.09$+$282112.86&SDSSJ002326.10$+$282112.81&5.85873&28.35357&17.81$\pm$0.01&19.29$\pm$0.14&18.34$\pm$0.01&1897.97&-1.57&0.13&QSO(0.24262$\pm$0.00003)\\
GNTJ002422.24$+$063051.18&SDSSJ002422.24$+$063051.16&6.09266&6.51422&18.98$\pm$0.02&20.17$\pm$0.18&19.42$\pm$0.02&1716.92&-2.00&0.85&\\
GNTJ002606.38$+$171223.03&SDSSJ002606.38$+$171223.02&6.52660&17.20640&18.16$\pm$0.01&20.01$\pm$0.07&19.39$\pm$0.18&1618.59&-2.69&1.38&\\
GNTJ002632.64$+$223812.21&SDSSJ002632.64$+$223812.23&6.63600&22.63673&19.01$\pm$0.03&20.80$\pm$0.15&20.30$\pm$0.05&1720.75&-1.06&0.87&\\
GNTJ002704.20$+$165209.38&SDSSJ002704.19$+$165209.41&6.76748&16.86927&18.92$\pm$0.01&20.31$\pm$0.15&19.77$\pm$0.10&1618.84&-1.44&0.59&\\
GNTJ002742.67$-$003858.23&SDSSJ002742.67$-$003858.19&6.92781&-0.64951&18.13$\pm$0.01&19.78$\pm$0.08&19.46$\pm$0.06&1734.74&-1.74&1.89&GALAXY(0.19383$\pm$0.00003)\\
GNTJ002820.77$+$302923.51&SDSSJ002820.76$+$302923.54&7.08653&30.48986&19.57$\pm$0.02&20.56$\pm$0.08&20.00$\pm$0.08&1737.94&-2.27&1.08&\\
GNTJ002941.57$+$113545.79&SDSSJ002941.57$+$113545.80&7.42320&11.59605&19.14$\pm$0.02&20.75$\pm$0.10&20.12$\pm$0.07&1900.49&-2.05&0.45&\\
\hline
\end{tabular}
\end{table}
\end{landscape}

\subsection{Nuclear transients published by {\it Gaia} Science Alerts}

\begin{figure}
\includegraphics[width=\columnwidth]{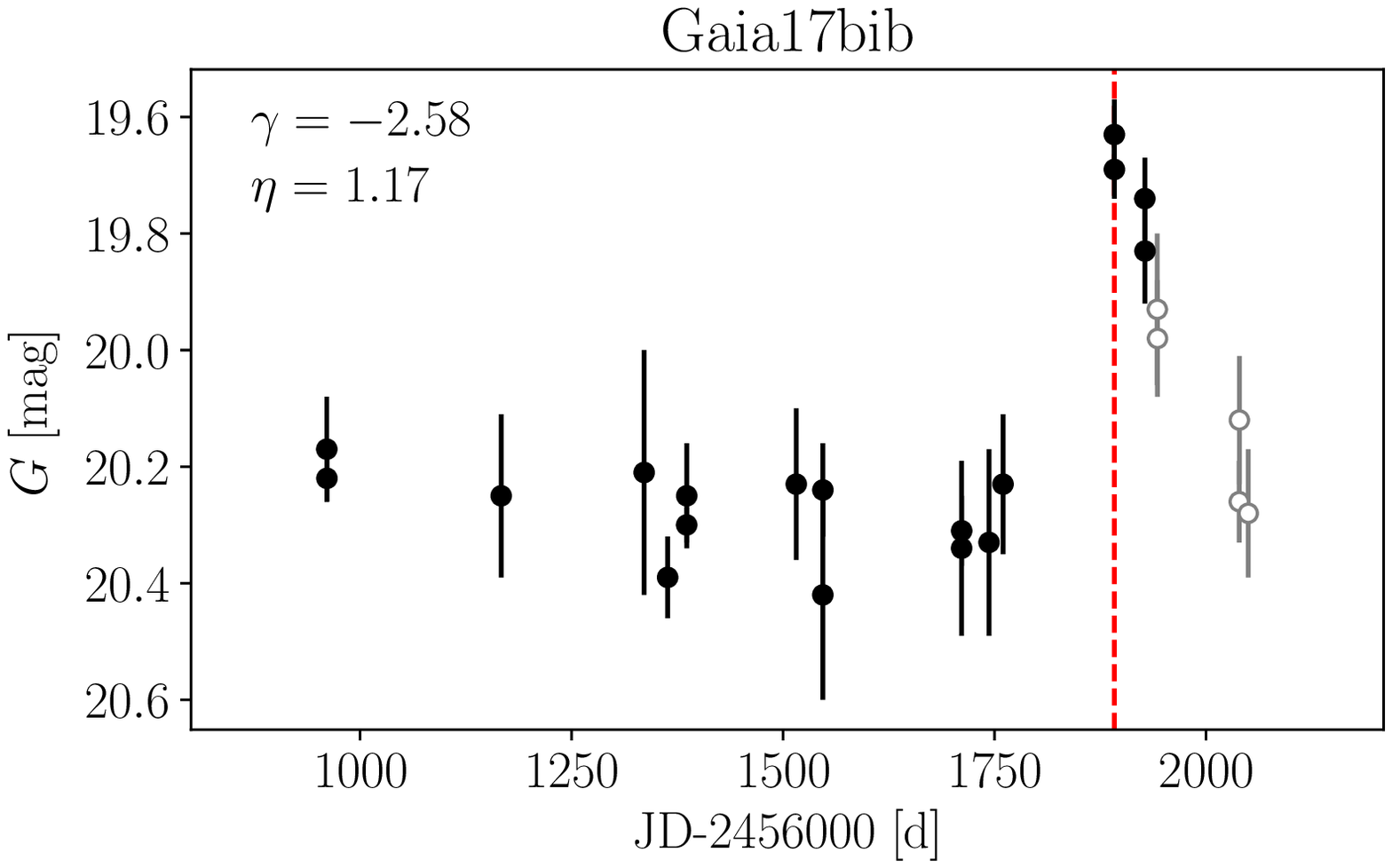}
\includegraphics[width=\columnwidth]{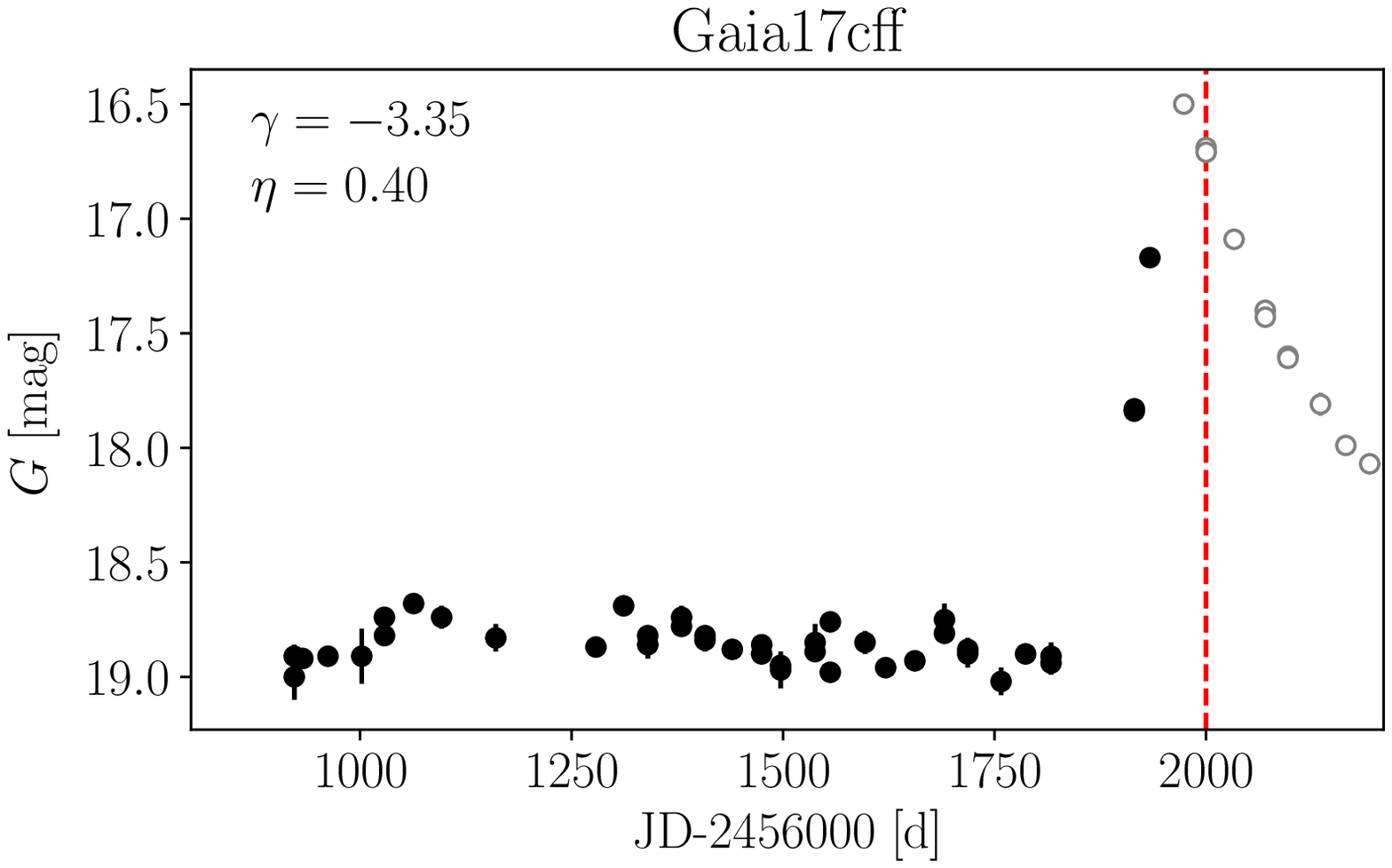}
\caption{Examples of light curves - known sources from {\it Gaia} Science Alerts re-discovered by this search. The red dashed line indicates the discovery time by GSA. The grey open circles denote data collected after our transient search window ended (June 2017). The transient Gaia17cff ({\it bottom}) was discovered by AlertPipe on 2017 September 3. The event was detected in our independent search two months before AlertPipe announced the discovery.}
\label{fig:lcknown}
\end{figure}

During the period of about 12 months between July 2016 and June 2017 the GSA team discovered about 50 transients in galaxy centres (transients associated with known galaxies and quasars within 0.5 arcsec) with tentative (uncalibrated) classifications as supernovae close to the host centre, or as quasar activity (see footnote 2). About half of them were found by the {\it NewSource} detector, and the other half by the {\it OldSource} detector.
Most of them (70 per cent) were detected in SDSS objects (photometrically and spectroscopically classified galaxies, spectroscopically classified quasars, and quasars candidates).
Additionally, in Tab.~\ref{tab:alrt} we show transients alerted by AlertPipe and re-discovered using our search on the skewness -- the reciprocal of von Neumann parameter space. 
The example light curve for the nuclear transients detected by the {\it OldSource} detector (Gaia17bib and Gaia17cff) with the discovery date is presented in Fig. \ref{fig:lcknown}. 
Two transients, Gaia17cff and Gaia17dko, were detected by AlertPipe in September 2017 and December 2017, respectively (so outside the period we consider for our independent search). Nevertheless, the transients are detected by our search using data collected until June 2017. I.e.~our search metric allowed the detection a few months earlier than the AlertPipe detection metric. The light curve of one of these transients is shown in Fig.~\ref{fig:lcknown} (bottom panel) where we indicate the data points that were not taken into account in the search described in this paper with grey open circles.

\begin{table*}
\centering
\caption{The {\it Gaia} Nuclear Transient (GNT) candidates previously alerted and published by {\it Gaia} Science Alerts pipeline. The source list provides the GSA name, the GNT source name, the SDSS galaxy ID, the alerting date by GSA, the JD at the time of the peak.}
\begin{tabular}{l l l l l}
\hline
GSA ID & GNT ID & SDSS galaxy & Alerting date & JD peak \\
\hline
Gaia16ajq & GNTJ145301.70$+$422127.82 & SDSSJ145301.70$+$422127.82 & 2016 Mar 29 & 2016 Dec 02 \\
Gaia17ays & GNTJ121112.76$+$381641.48 & SDSSJ121112.76$+$381641.48 & 2017 Apr 08  & 2017 Apr 08 \\
Gaia17bib & GNTJ100443.32$-$022427.35 & SDSSJ100443.32$-$022427.32 & 2017 May 18 & 2017 May 18 \\
Gaia17cff & GNTJ171955.85$+$414049.46 & SDSSJ171955.85$+$414049.45 & 2017 Sep 03 & 2017 Jun 10 \\
Gaia17dko & GNTJ124027.76$-$051400.77 & SDSSJ124027.76$-$051400.71 & 2017 Dec 26 & 2017 May 28 \\
\hline
\end{tabular}
\label{tab:alrt}
\end{table*}

\subsection{Examples of known nuclear transients missed by {\it Gaia} Science Alerts}

There are various reasons why transients discovered in other surveys may not be discovered or alerted on by AlertPipe. The main reason lies in the AlertPipe settings (see Hodgkin et al. in prep.). For example, the tidal disruption event discovered by the intermediate Palomar Transient Factory (iPTF) survey at 66.6 Mpc in the centre of galaxy Mrk950 (SDSSJ002957.05$+$325337.2) (\citealt{2017ApJ...844...46B}) was not found by AlertPipe. The reason for this is that iPTF16fnl was only detected on one of the two field of views of {\it Gaia}, the second field of view did not pass over the source 106.5 minutes later due to the satellite precession. The next {\it Gaia} observation was taken after the transient declined (95 days later). 
In order for a transient to be found by AlertPipe it must be detected in each of the {\it Gaia} field of views within 40 days of each other.
This condition is not fulfilled here, hence the transient was missed by AlertPipe even though it was relatively bright. 

\begin{figure}
\includegraphics[width=\columnwidth]{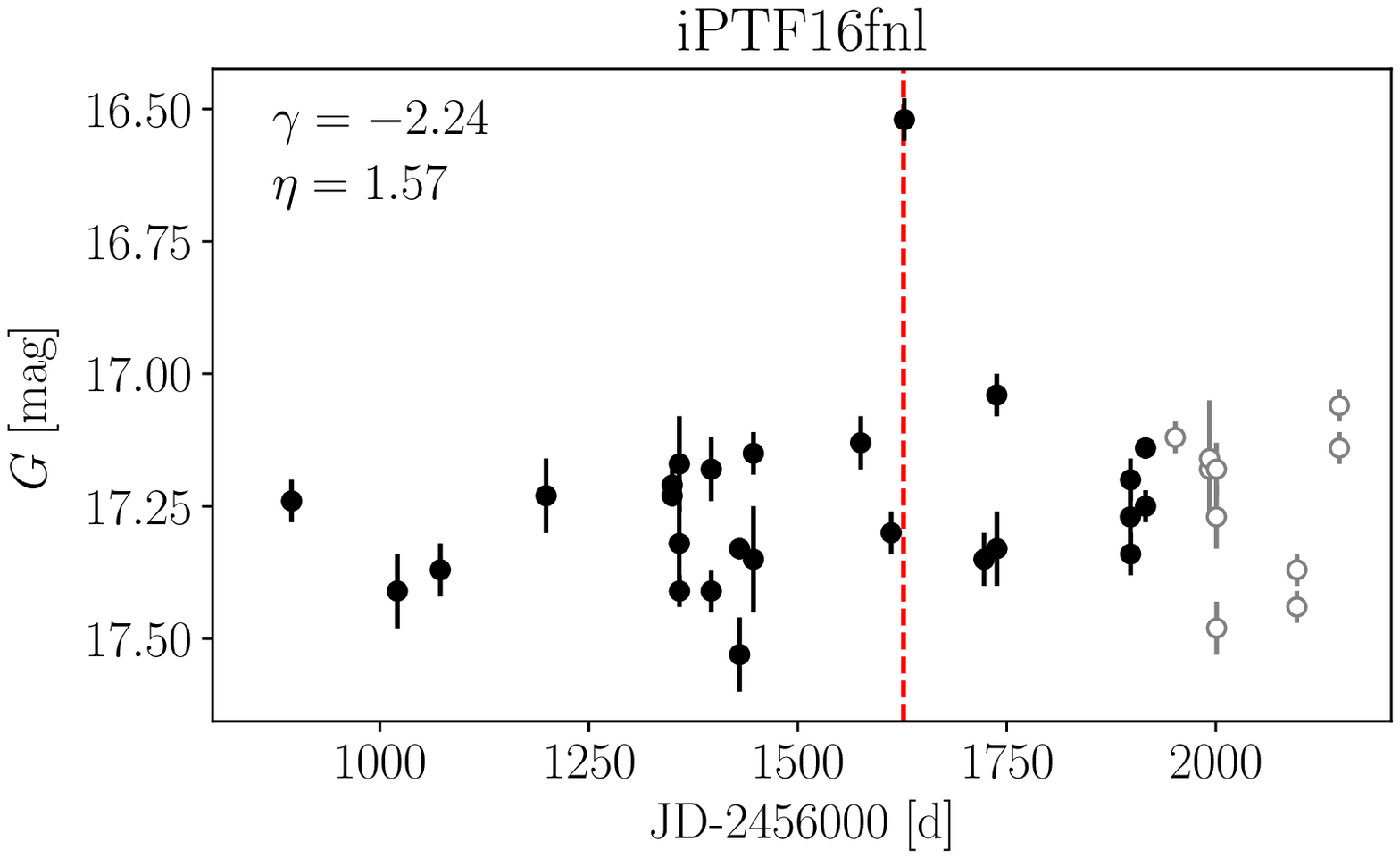}
\includegraphics[width=\columnwidth]{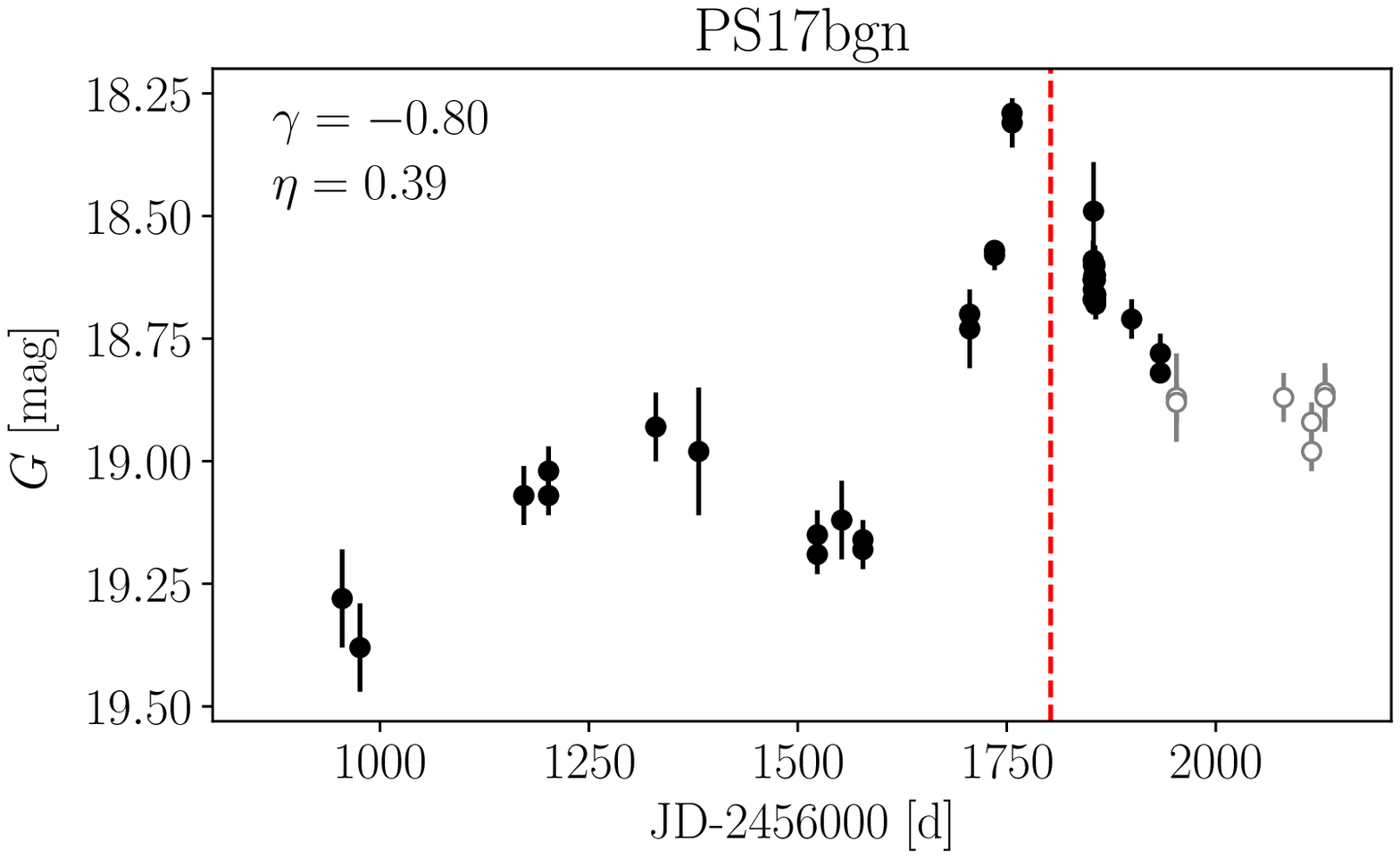}
\caption{Examples of light curves of transient sources discovered by other surveys but not by {\it Gaia} even though the sources were detected. The dashed red line indicates the discovery time. {\it Top:} The tidal disruption event iPTF16fnl detected by the iPTF survey (\citealt{2017ApJ...844...46B}) on 2016 August 29. {\it Gaia} observed the transient on 2016 August 26.  {\it Bottom:} A candidate superluminous supernova from the PanSTARRS survey - PS17bgn (\citealt{2017ATel10105....1T}) was discovered on 2017 February 18. The maximum of {\it Gaia} light curve was on 2017 January 2. In both cases {\it Gaia} observed the transients before they were detected by the other surveys.}
\label{fig:lcfnlbgn}
\end{figure}

Another transient, the source PS17bgn (SN2017bcc), was detected by the Panoramic Survey Telescope and Rapid Response System (PanSTARRS). It falls in the centre of the galaxy SDSSJ113152.97$+$295944.8 (\citealt{2017ATel10105....1T}). Here, the host showed previous variability with an amplitude of 0.4 mag. Also this transient was missed by AlertPipe as the outburst was not bright enough, the {\it Gaia} data point in outburst did not stand-out sufficiently compared to the detected baseline and its variability. The {\it Gaia} light curves for both transients are presented in Fig. \ref{fig:lcfnlbgn}.

These two examples show that the previous variability and the requirement on AlertPipe for a very low number of false-positives complicates the detection of transients in galaxy centres with the {\it OldSource} detector. However, these transients were picked out by our independent search on the skewness -- the reciprocal of von Neumann parameter plane. 

%Spectroscopic verification
%GNT J233855.86+433916.86 1925673944649921920 1
%GNT J232841.41+224847.96 2839665181628793984 2
%GNT J002326.09+282112.86 2857092544007792896 N 

\subsection{Spectroscopic verification}

\begin{figure*}
\begin{tabular}{ccc}
\includegraphics[width=80mm]{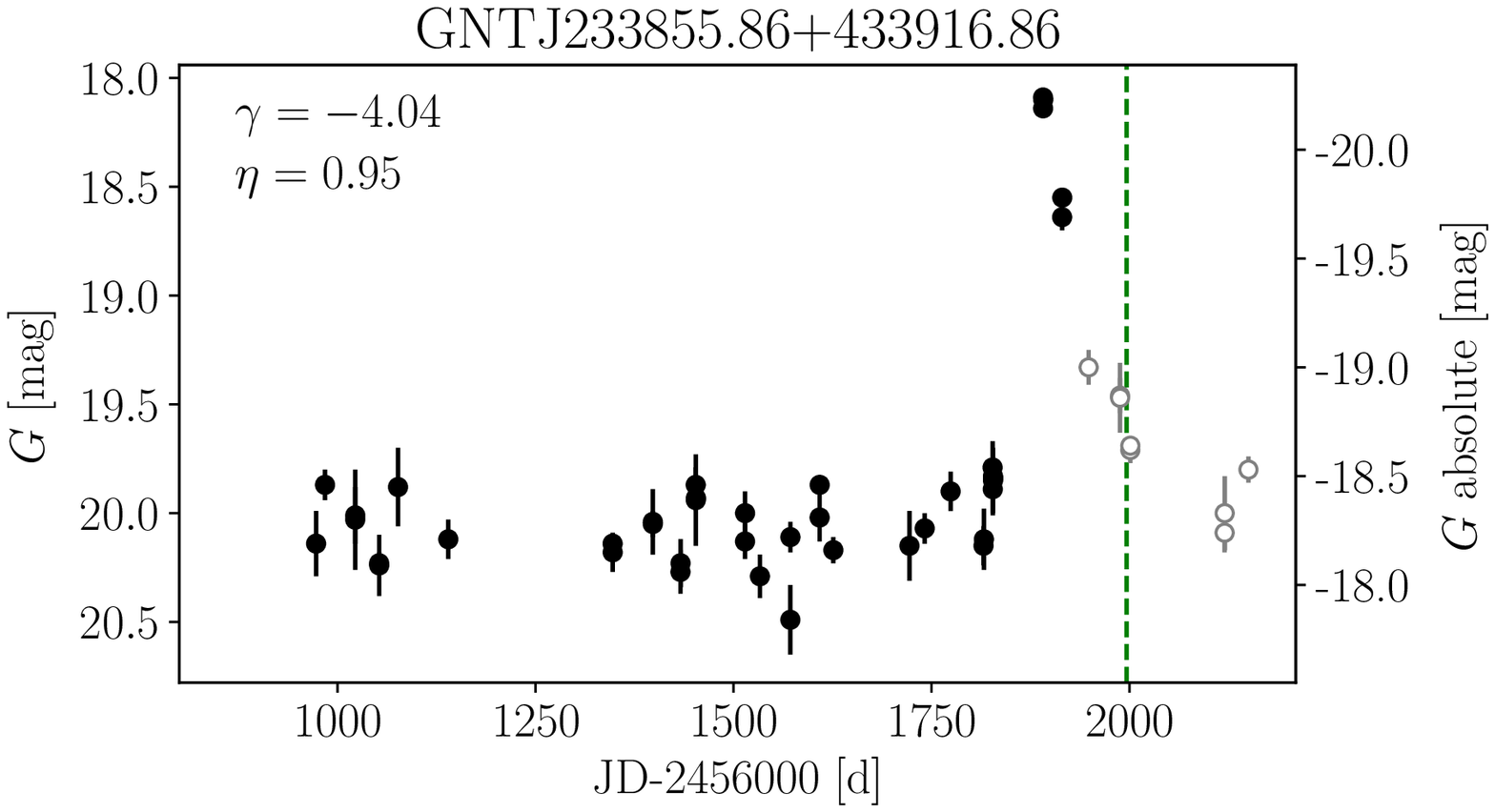} & 
\includegraphics[width=80mm]{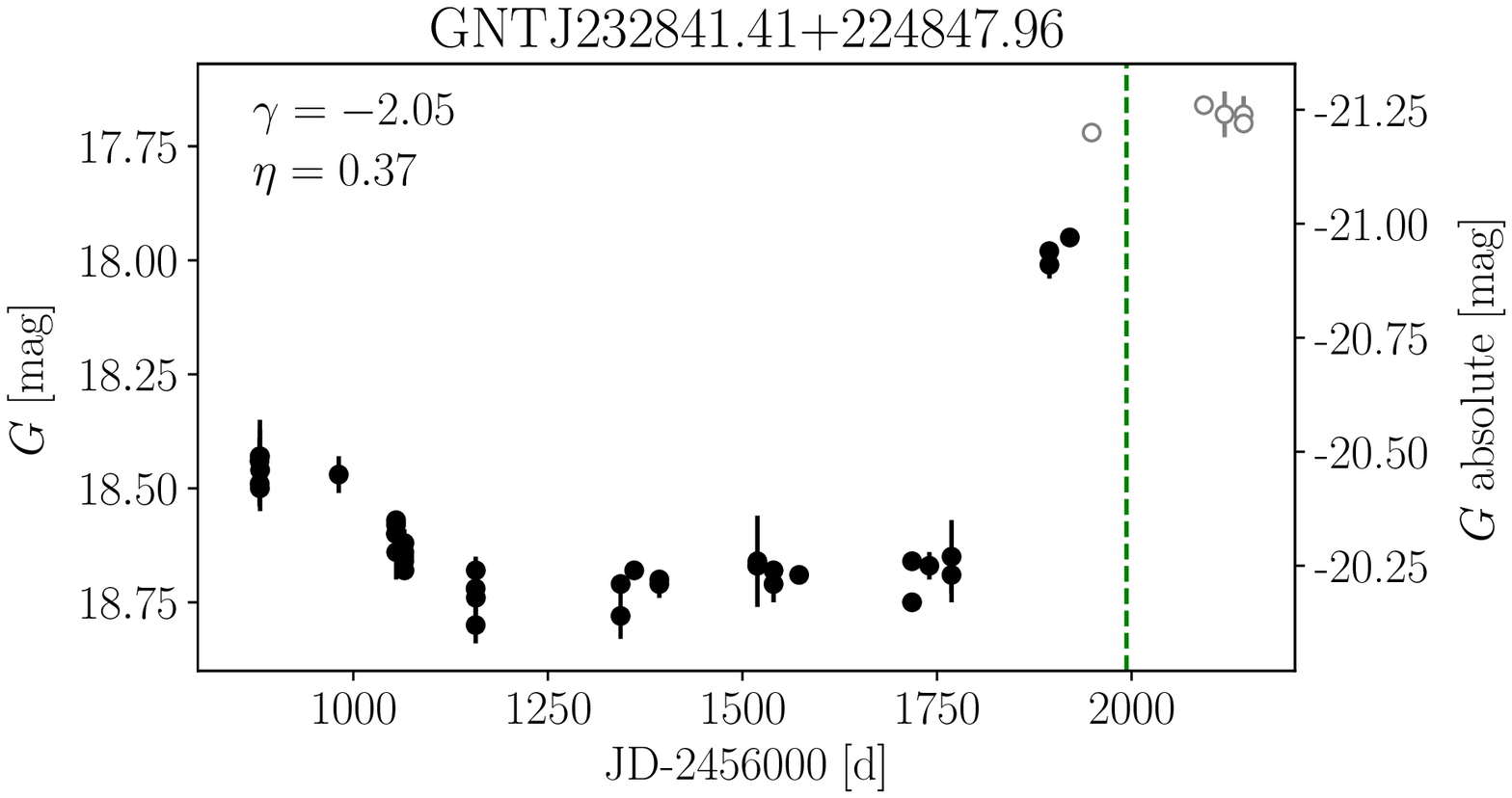} \\
\includegraphics[scale=1.0]{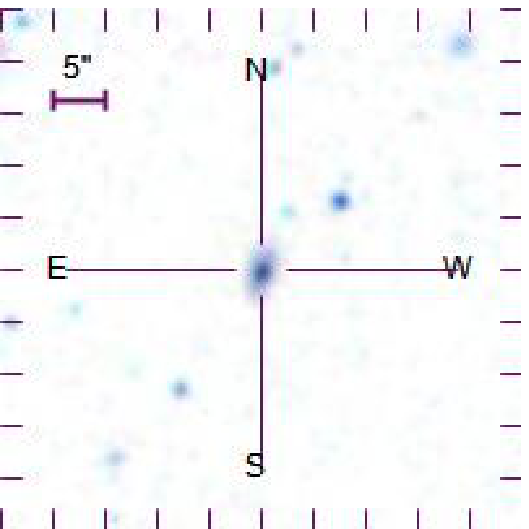}  & \includegraphics[scale=1.0]{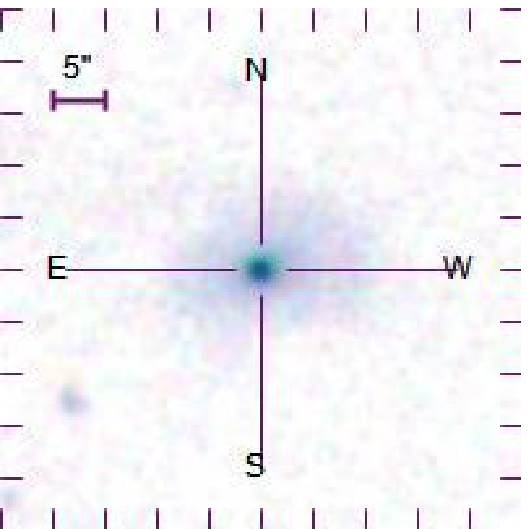}  \\
\includegraphics[height=50mm]{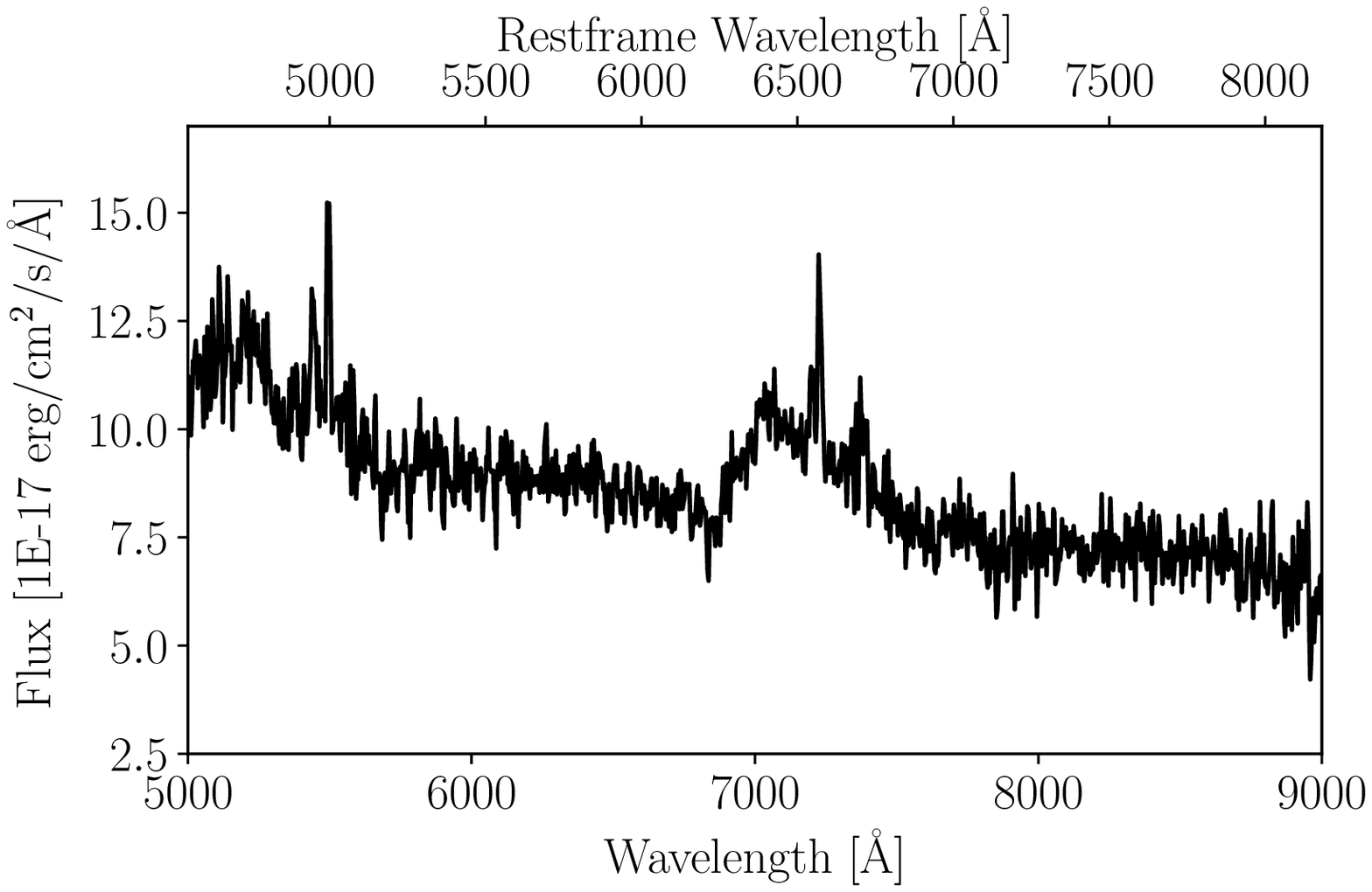} & \includegraphics[height=50mm]{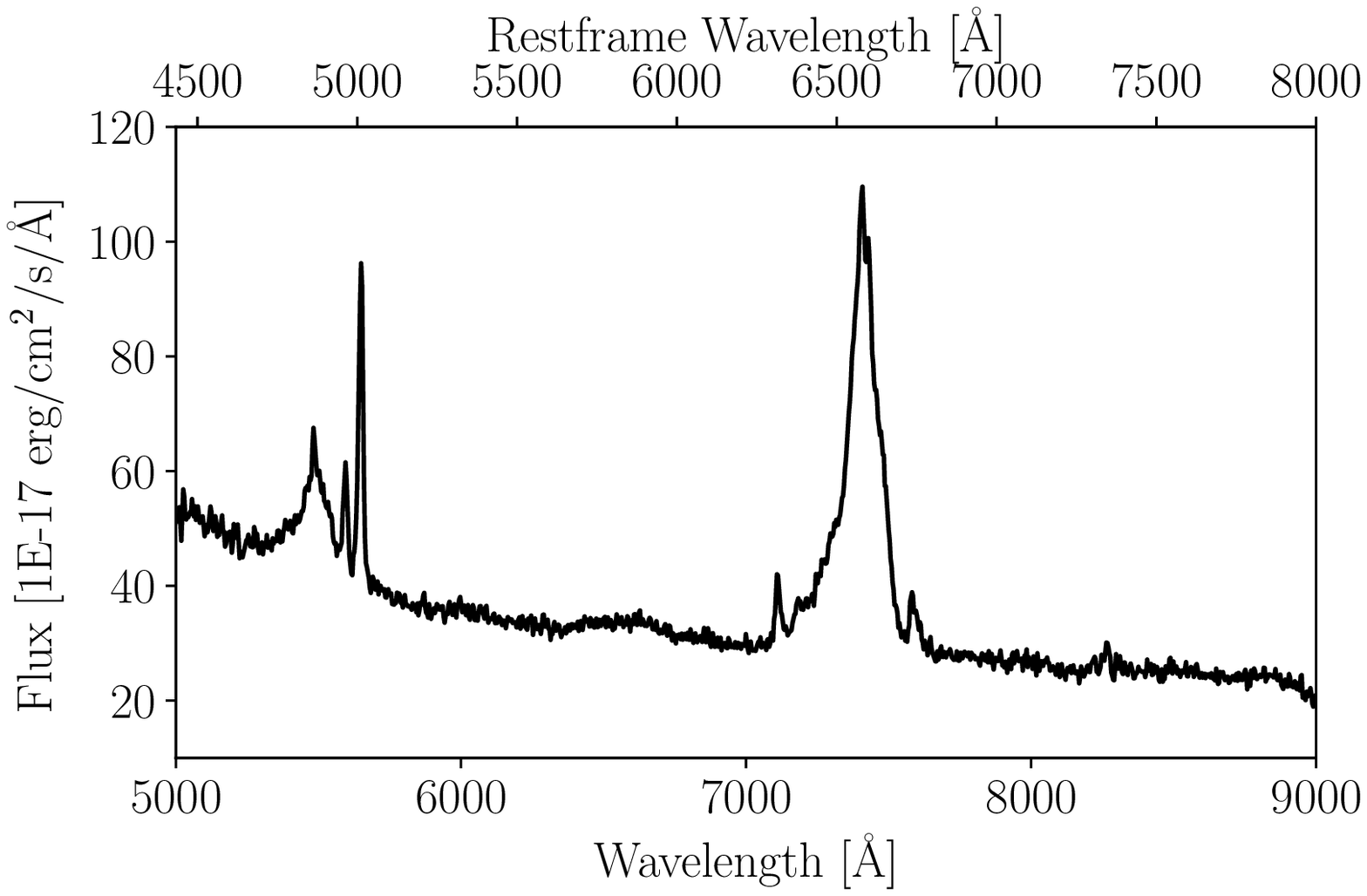} \\
\end{tabular}
\caption{Examples of light curves ({\it top panels}) of candidate {\it Gaia} nuclear transients with SDSS finding charts ({\it middle panels}; $50\arcsec\times50\arcsec$) and classification spectra ({\it bottom panels}). The spectra were taken in August 2017 (green dashed line in light curve plots), at least 2 months after the transients were first detected in the {\it Gaia} data. Telluric features are corrected. {\it Left:} GNTJ233855.86$+$433916.86 transient in the SDSS galaxy with two {\it Gaia} sources. The object was split into two sources during the Initial Data Treatment. The merged light curve was analysed and presented in the top panel. The grey open circles denote {\it Gaia} data collected after the period that we employed for our transient search. The galaxy was inactive for at least last two years. The recent flare of about 2 mag lasted for about 200 days. We obtained a redshift $\sim$0.10. The broad and narrow \ion{H}{$\alpha$} emission lines are visible. The broad component of the \ion{H}{$\alpha$} emission line likely covers expected \ion{N}{ii}, \ion{O}{i}, and \ion{S}{ii} emission lines. {\it Right:} In the light curve of transient GNTJ232841.41$+$224847.96 we notice a rise of about 1 mag above the baseline, moreover previous variability is visible. The spectrum is similar to that of a broadline quasar at redshift $\sim$0.129. Several emission lines (\ion{H}{$\alpha$}+\ion{N}{ii}, \ion{O}{i}, \ion{S}{ii}, \ion{H}{$\beta$}, \ion{O}{iii})  were detected. Note: The plots, light curves and spectra for these objects are available in ascii format from the online journal.}
\label{fig:lcspec}
\end{figure*}

\begin{figure*}
\includegraphics[width=130mm]{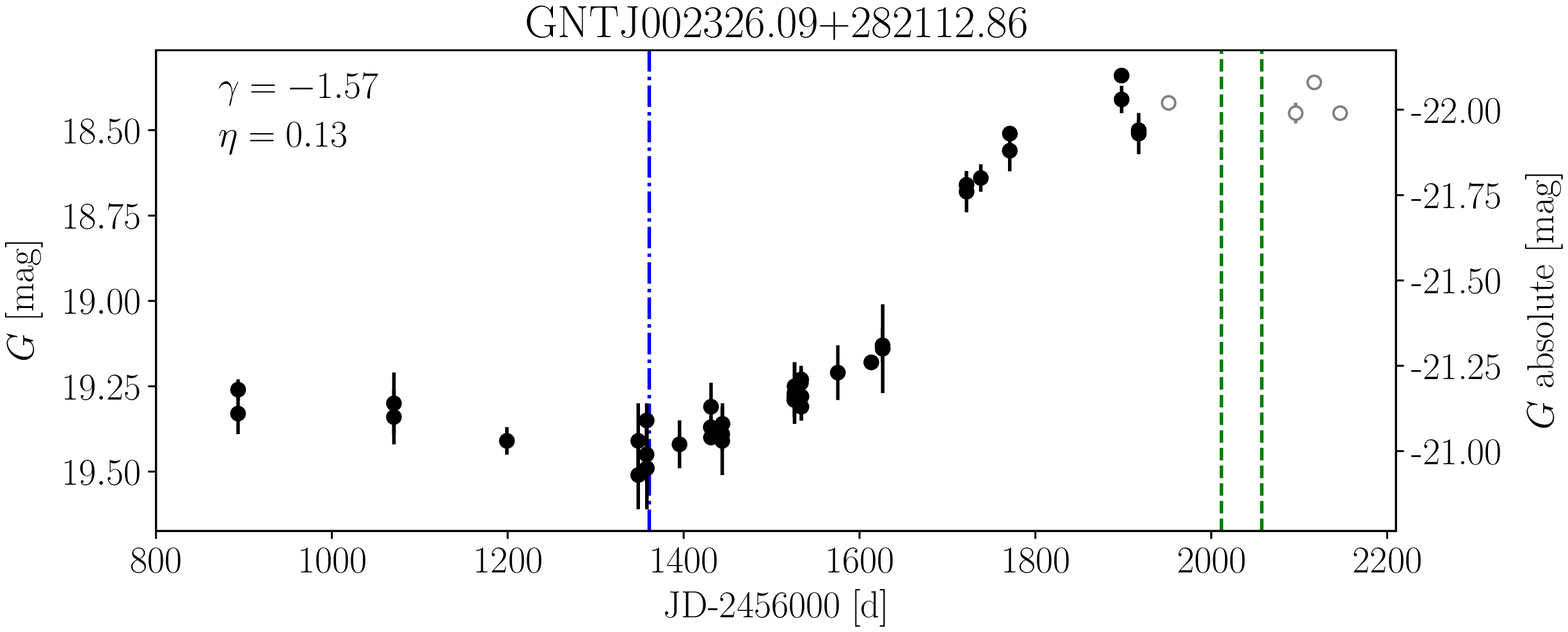}
\includegraphics[scale=0.8]{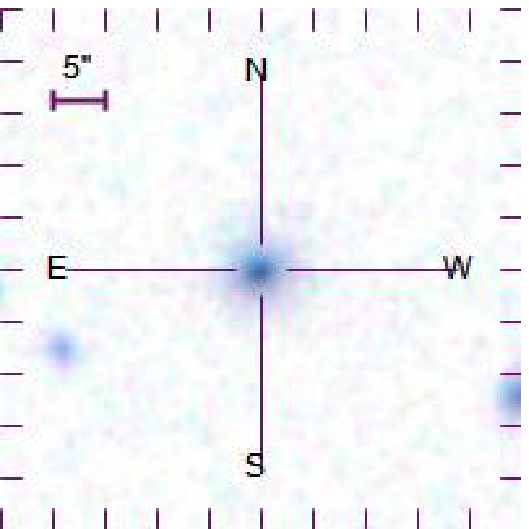}
\caption{Candidate nuclear transient GNTJ002326.094$+$282112.86. {\it Left:} The {\it Gaia} light curve analysed during the search (black full circles). The grey open circles denote data collected after our transient search window ended (June 2017 -- these points were not taken into account during the analysis). The blue dot-dashed line indicates the time when the host spectrum was taken by SDSS (December 2015). The green dashed lines indicate the time of the WHT spectroscopic observations in September and October 2017. {\it Right:} The SDSS finding chart ($50\arcsec\times50\arcsec$). Note: The light curve for this object is available in ascii format from the online journal.}
\label{fig:gaiaNlc}
\end{figure*}

\begin{figure*} 
\includegraphics[width=150mm]{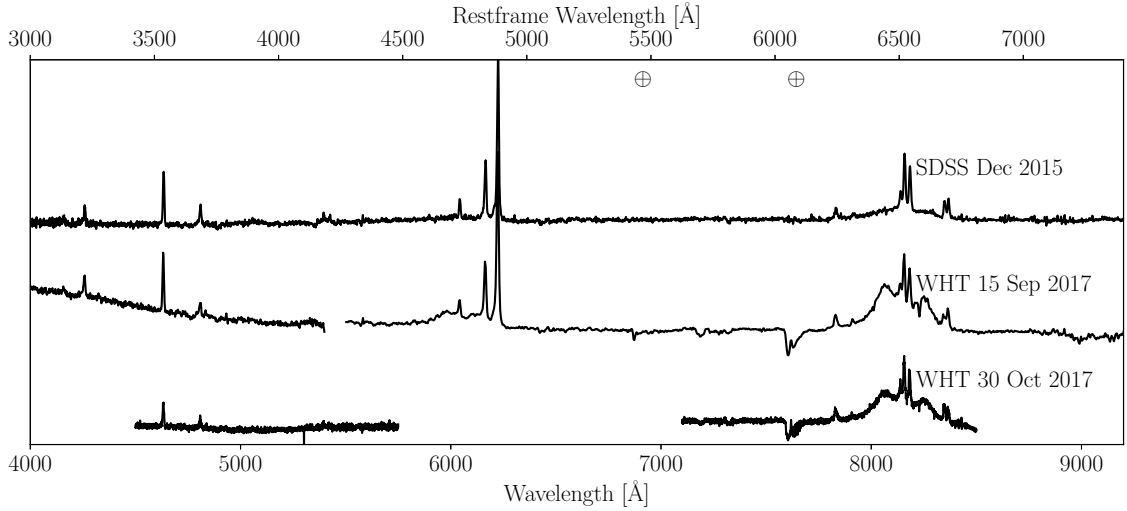}
\caption{Spectra of the candidate nuclear transient GNTJ002326.09$+$282112.86. Top - an archival spectrum from SDSS DR14 obtained in December 2015. The object was classified by SDSS as a broadline QSO starburst at redshift $0.24262\pm0.00003$. Two spectra to classify the outburst event were taken with the WHT on 2017 September 14 and on 2017 October 30 during outburst. During the outburst a very broad emission component appeared around the \ion{H}{$\alpha$} line. The Y-axis of the spectra in this plot is the flux $+$ constant value.}
\label{fig:gaiaNspec}
\end{figure*}

For three objects with the most recent transients we took classification spectra in late August/mid- September. The observations were performed with the 4.2m William Herschel Telescope (WHT) located on La Palma, Spain.
An overview of the spectroscopic observations is presented in Tab. \ref{tab:spectra}.
The {\it Gaia} light curves, finding charts, and WHT spectra for three objects are presented in Fig.~\ref{fig:lcspec}, \ref{fig:gaiaNlc}, and \ref{fig:gaiaNspec}, respectively.
The spectra were reduced with the standard steps such as a bias level subtraction, a flat-field correction, and a wavelength and flux calibration using \textsc{iraf}. Cosmic rays were removed using the {\it lacosmic} package (\citealt{2001PASP..113.1420V}). The typical root mean square deviation of the applied wavelength solution is $<0.2$\AA.
These classification spectra are also available in ascii file format from the online journal.

\begin{table*}
\centering
\caption{An overview of optical spectroscopy of the targets and instrumental set-ups used to classify the most recent nuclear transient candidates. All spectra were taken with William Herschel Telescope. We provide the grating, the slit width and the seeing, the approximate wavelength coverage ($\lambda$), the exposure time $\mathrm{T_{exp}}$ in seconds, and the number of exposures $N$. The reduced spectra are available from the online journal in ascii format.}
\begin{tabular}{l l l l l l l}
\hline
GNT ID & Date & Instrument & Grating & Slit('')/seeing('') & $\lambda$(\AA) & $\mathrm{T_{exp}}$ (s)  $\times$ $N$ \\
\hline
GNTJ233855.86$+$433916.86 & 2017 Aug 27 & ACAM & V400 & 1.0/0.7-1.5 & 5000-9000 & 900 $\times$ 2\\
 & 2017 Aug 28 & ACAM & V400 & 1.0/0.8-1.5 & 5000-9000 & 900 $\times$ 4\\
 & 2017 Aug 27 & ISIS & R600B,R600R & 1.0/0.7-1.5 & 3800-7300 & 1800 $\times$ 1\\
 & 2017 Aug 29 & ISIS & R300B,R316R & 1.0/1.5-2.0 & 3000-8000 & 1800 $\times$ 2\\
GNTJ232841.41$+$224847.96 & 2017 Aug 27 & ACAM & V400 & 1.0/0.7-1.5 & 5000-9000 & 600 $\times$ 1 \\
GNTJ002326.09$+$282112.86 & 2017 Sep 14 & ISIS & R300B,R158R & 1.0/1.0 & 3500-9000 & 1800 $\times$ 1\\
 & 2017 Sep 15 & ISIS & R300B,R158R & 1.0/1.0 & 3500-9000 & 1800 $\times$ 1 \\
 & 2017 Oct 29 & ISIS & R600B & 1.0/0.7-1.0 & 4400-5800 & 1800 $\times$ 3\\
 & 2017 Oct 30 & ISIS & R600R & 1.0/0.7 & 7000-8500 & 1800 $\times$ 3\\
\hline
\end{tabular}
\label{tab:spectra}
\end{table*}

%.........................
 
For the SDSS galaxy SDSSJ233855.86$+$433916.87 we found the counterpart source in the GSA DB (GNTJ233855.86$+$433916.86) which is a candidate nuclear transient on the skewness -- the reciprocal of von Neumann parameter plane. The object was split into two separate sources due to an ambiguous cross-match during the Initial Data Treatment (for the further discussion on data cross-matching see \citealt{2016A&A...595A...3F,2017A&A...599A..50A}) and hence the transient was not found by AlertPipe. In Fig. \ref{fig:lcspec} (left panels) we show the light curve, the SDSS finding chart and the WHT spectrum taken at a late phase in the transient light curve. The galaxy was inactive for at least two years. The recent flare of about 2 mag at maximum lasted for about 200 days. We obtained a redshift $\sim$0.10 from the WHT spectrum. The spectrum shows both a broad and a narrow \ion{H}{$\alpha$} emission line.

In Fig. \ref{fig:lcspec} (right panels) we show the light curve, the SDSS finding chart, and the WHT spectrum for candidate GNTJ232841.41$+$224847.96 in the galaxy SDSSJ232841.40$+$224848.02. 
In the light curve we notice a rise of about 1 magnitude above the baseline, moreover previous variability is visible but on a much lower scale. The transient might be still active as it was still rising. The spectrum is similar to a broadline quasar at redshift $\sim$0.129. Several emission lines (\ion{H}{$\alpha$}+\ion{N}{ii}, \ion{O}{i}, \ion{S}{ii}, \ion{H}{$\beta$}, \ion{O}{iii}) were detected.

\begin{figure} 

\includegraphics[width=\columnwidth]{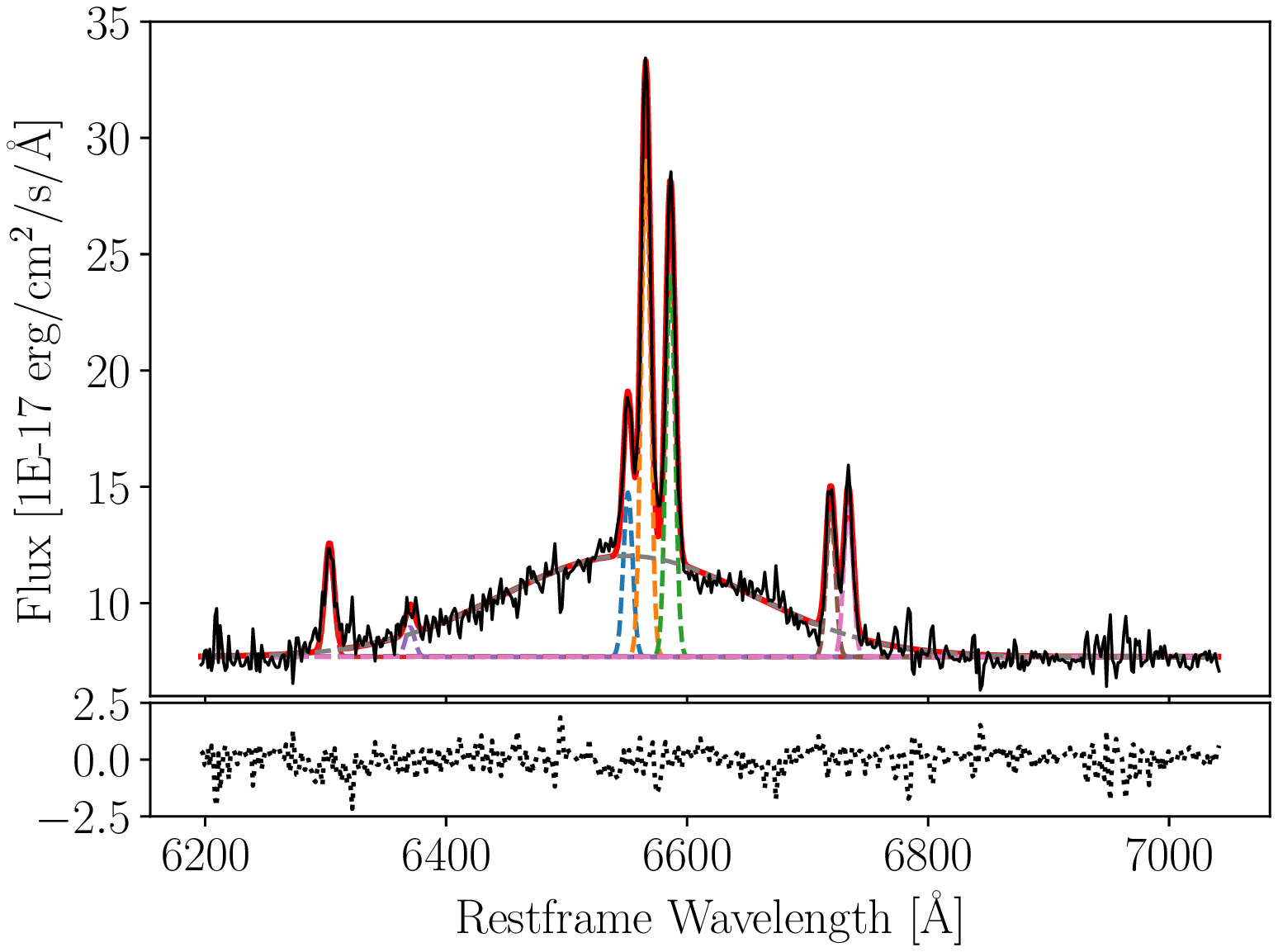}
\includegraphics[width=\columnwidth]{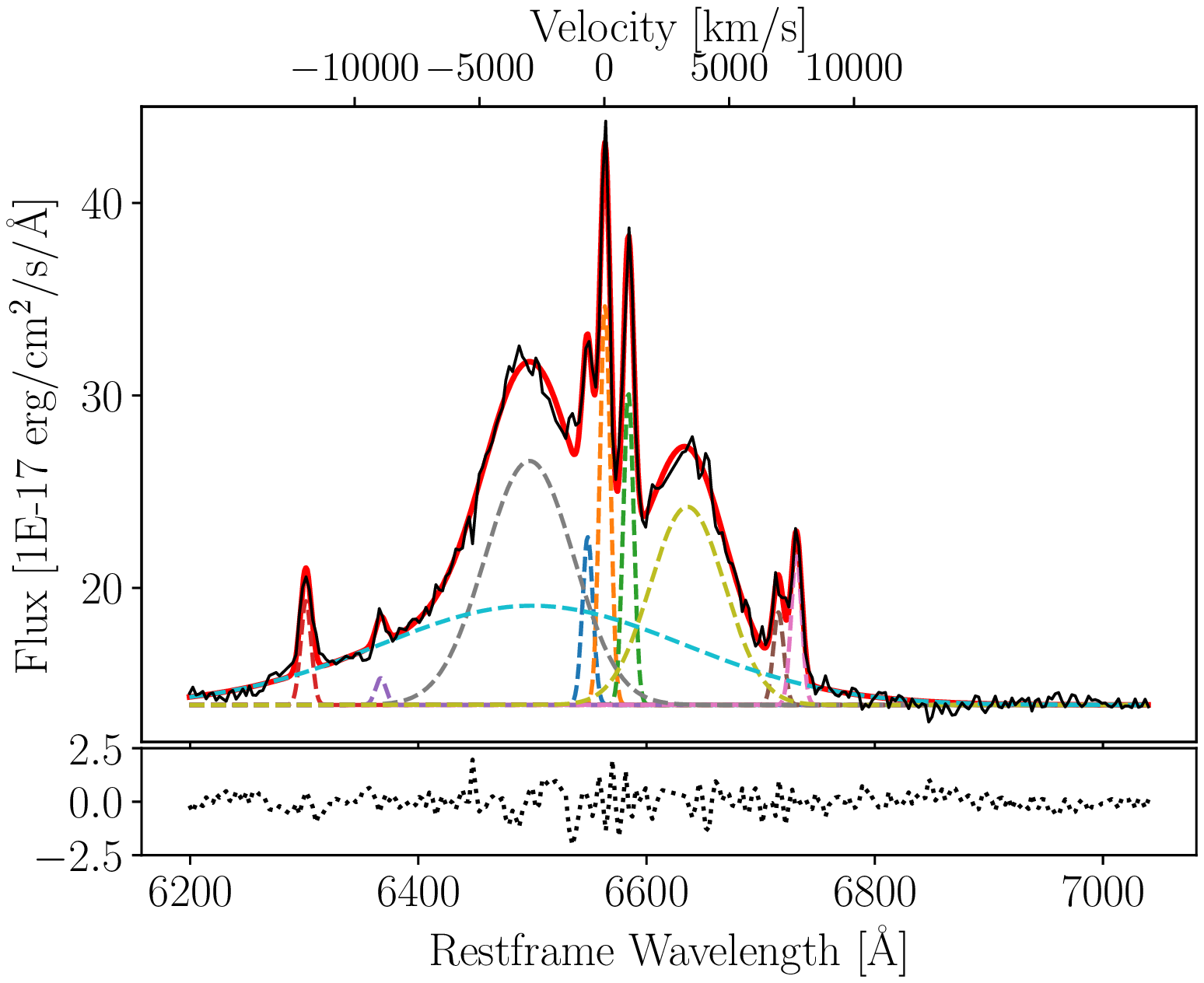}
\caption{Candidate nuclear transient GNTJ002326.09$+$282112.86. {\it Top:} The complex broad \ion{H}{$\alpha$} region in the SDSS spectrum taken in December 2015. Red line - the best fit, black line - the data, dashed lines - fit components, bottom dotted line - residuals. We fitted the narrow emission lines \ion{O}{i}, \ion{N}{ii}, \ion{S}{ii}, \ion{H}{$\alpha$}, and one broad component around \ion{H}{$\alpha$} with Gaussians. {\it Bottom:} The complex broad \ion{H}{$\alpha$} region in the WHT spectrum taken in September 2017. Red line - the best fit, black line - the data, dashed lines - fit components, bottom dotted line - residuals. We fitted the narrow emission lines \ion{O}{i}, \ion{N}{ii}, \ion{S}{ii}, \ion{H}{$\alpha$}, and three broad components around \ion{H}{$\alpha$} with Gaussians. The broad component was previously present, however it changed and two new red- and blue-shifted parts appeared. The summary of fit results is presented in Tab. \ref{tab:gaiaNspecHa}. The \ion{H}{$\alpha$} region is also presented in velocity space.}
\label{fig:gaiaNspecHa}
\end{figure}

%specN 
The light curve of GNTJ002326.09$+$282112.86 (Fig. \ref{fig:gaiaNlc}, left panel) shows an outburst of 1 magnitude that started mid-2016 in the galaxy SDSSJ002326.10$+$282112.81. During 200 days the transient rose up to $G=18.35$ mag. The absolute magnitude at maximum is $G=-22.14$ mag. The spectrum of the host is available from SDSS DR14 where the object was classified as a broadline QSO starburst at redshift $0.24262\pm0.00003$. This spectrum was taken in December 2015 well before the outburst started. We took spectra with the WHT on 2017 September 14 and on 2017 October 30. All three spectra are presented in Fig.~\ref{fig:gaiaNspec}, left panel. During the transient event a broad red- and blue-shifted \ion{H}{$\alpha$} emission line appeared. The detailed fit of the complex \ion{H}{$\alpha$} wavelength region is presented in Fig. \ref{fig:gaiaNspecHa} and in Tab. \ref{tab:gaiaNspecHa}. Several narrow emission lines and three broad components were fitted with Gaussian functions. The broad \ion{H}{$\alpha$} component is presented in velocity space in Fig. \ref{fig:gaiaNspecHa} (bottom panel). The flux of broad components peaks around $+/-3500$ km/s indicating an outflow driven by for instance a wind.


\begin{table*}
\centering
\caption{A summary of fits of Gaussian to the \ion{H}{$\alpha$} region in the spectrum of the candidate nuclear transient GNTJ002326.09$+$282112.86 taken in September 2017 and in the archival spectrum taken in December 2015. The spectra were fitted in the rest frame using redshift 0.24262. The apparent difference in FWHM for the narrow lines is caused by the different resolution of the spectra and not due to intrinsic widening.}
\begin{tabular}{l l l l l l l}
\hline
& \multicolumn{3}{c}{WHT Sep 2017} & \multicolumn{3}{c}{SDSS Dec 2015} \\
Line & Wavelength [\AA] & FWHM [\AA]  & EW [\AA] & Wavelength [\AA] & FWHM [\AA]  & EW [\AA] \\
\hline
\ion{H}{$\alpha$} & 6563.65 $\pm$ 0.11 & 10.89 $\pm$ 0.17 & 17.25 $\pm$ 0.44 & 6565.71 $\pm$ 0.06 & 8.87 $\pm$ 0.11 & 26.17 $\pm$ 0.49 \\
\ion{N}{ii} & 6548.43 $\pm$ 0.26 & 10.89 $\pm$ 0.17 & 7.26 $\pm$ 0.36 & 6550.91 $\pm$ 0.19 & 8.87 $\pm$ 0.11 & 8.68 $\pm$ 0.35 \\
\ion{N}{ii} & 6584.11 $\pm$ 0.13 & 10.89 $\pm$ 0.17 & 13.48 $\pm$ 0.40  & 6586.19 $\pm$ 0.08 & 8.87 $\pm$ 0.11 & 20.24 $\pm$ 0.43 \\
\ion{O}{i} & 6301.49 $\pm$ 0.37 & 10.89 $\pm$ 0.17 & 4.49 $\pm$ 0.29 & 6303.18 $\pm$ 0.28 & 8.87 $\pm$ 0.11 & 5.70 $\pm$ 0.32 \\
\ion{O}{i} & 6367.09 $\pm$ 1.45 & 10.89 $\pm$ 0.17 & 1.16 $\pm$ 0.29 & 6370.13 $\pm$ 1.04 & 8.87 $\pm$ 0.11 & 1.56 $\pm$ 0.32 \\
\ion{S}{ii} & 6715.59 $\pm$ 0.44 & 10.89 $\pm$ 0.17 & 4.01 $\pm$ 0.30 & 6719.23 $\pm$ 0.22 & 8.87 $\pm$ 0.11 & 7.62 $\pm$ 0.35 \\
\ion{S}{ii} & 6731.33 $\pm$ 0.28 & 10.89 $\pm$ 0.17 & 6.43 $\pm$ 0.32 & 6733.75 $\pm$ 0.21 & 8.87 $\pm$ 0.11 & 7.91 $\pm$ 0.35\\
\ion{H}{$\alpha$}-broad & 6499.76 $\pm$ 6.22 & 314.87 $\pm$ 13.98 & 124.15 $\pm$ 11.97 & 6550.27 $\pm$ 1.73 & 245.82 $\pm$ 5.54 & 147.49 $\pm$ 4.30 \\
\ion{H}{$\alpha$}-blue & 6497.39 $\pm$ 0.67 & 91.52 $\pm$ 2.82 & 88.71 $\pm$ 4.11 & & &\\
\ion{H}{$\alpha$}-red & 6635.86 $\pm$ 0.71 & 76.67 $\pm$ 2.31 & 60.37 $\pm$ 2.73 & & & \\

\hline
\end{tabular}
\label{tab:gaiaNspecHa}
\end{table*}

The spectra and light curves show no similarity with known Type Ia and core collapse supernova spectra and light curves as verified using supernova spectral templates through the SNID tool (\citealt{2007ApJ...666.1024B}).
However, as the broad \ion{He}{II} line that is considered a typical TDE feature, has been found to vary in time and between TDEs (e.g. \citealt{2014ApJ...793...38A}) our single epoch spectroscopy does not provide sufficient evidence to differentiate between a TDE scenario or peculiar AGN variability (such as changing look quasars) for candidate transients GNTJ233855.86$+$433916.86 and GNTJ232841.41$+$224847.96.
The candidate transient GNTJ002326.09$+$282112.86 has a very broad emission component around the \ion{H}{$\alpha$} line and no indication of \ion{He}{II} lines. And given that for this source we have three epochs of spectroscopy we deem it likely that this candidate transient is due to peculiar AGN activity.

\section{Discussion}

\cite{2016MNRAS.455..603B} predicted that {\it Gaia} Science Alerts will discover about 215 nuclear transients (from supernovae and TDEs) per year brighter than 19 mag and with an increase in magnitude of 0.3 mag or more. These sources would be discovered using both the {\it NewSource} and {\it OldSource} detectors. Our study presented here comprises about one third of the sky, and we report $\sim$160 ($\sim$480) candidates for transients brighter than 19 mag (20.5 mag). All these transients were discovered using historical data from the GSA DB (the same used by the {\it OldSource} detector). Our sample does not contain transients that would have been found by the {\it NewSource} detector. 
Our dedicated search for nuclear transients may be more sensitive than AlertPipe that is designed to discover all types of transients (like supernovae, cataclysmic variables, microlensing events, flare stars etc) with a low false-positive rate. Nevertheless, significant manual vetting of candidate nuclear transients has been necessary.

\begin{figure}
\includegraphics[width=\columnwidth]{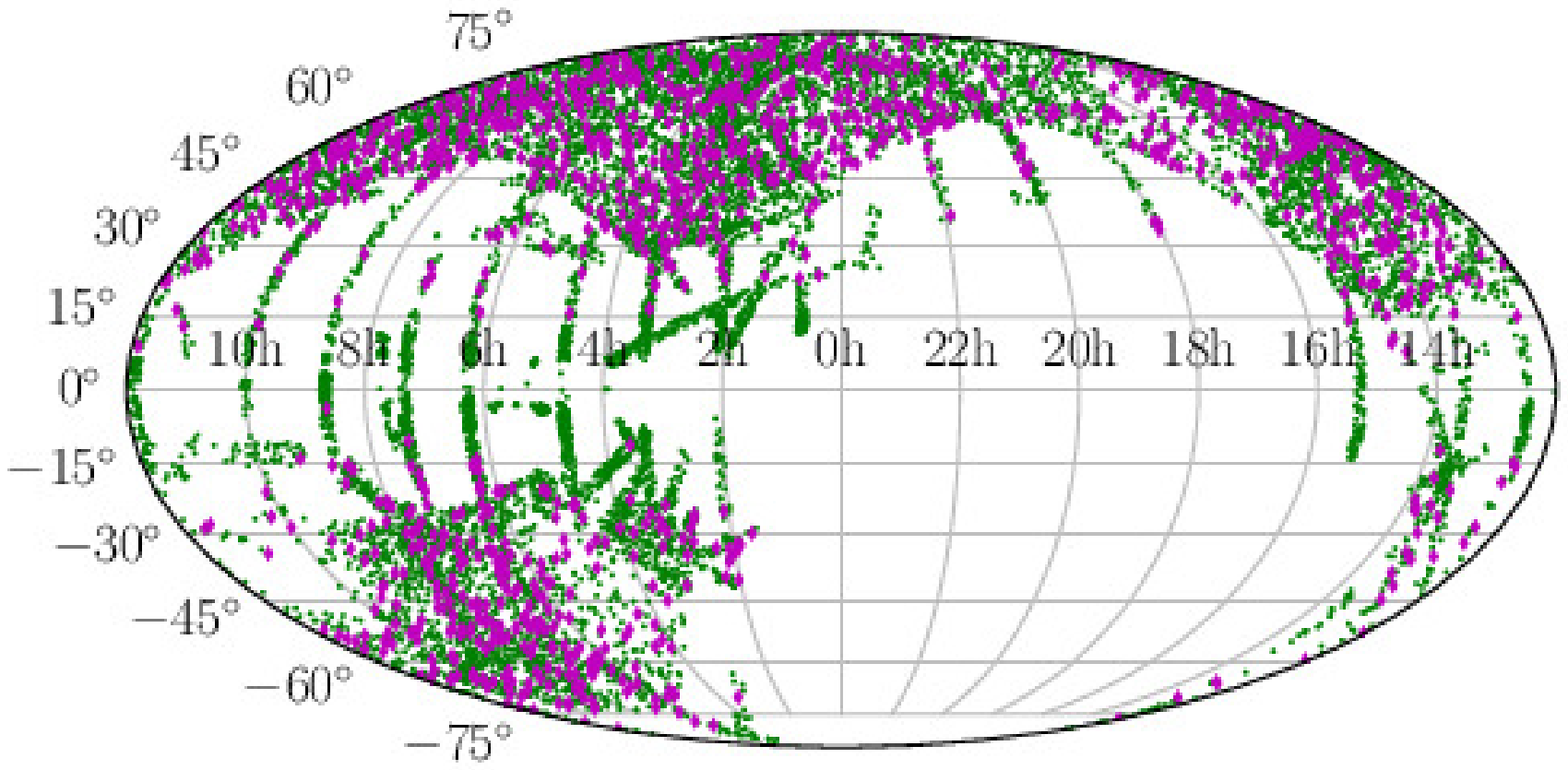}
\caption{The map of all our candidate transients in Galactic coordinates, i.e. the 6k candidates (green dots), before our eye-balling reduced this to $\sim$480 (magenta diamonds). Most of the real transients were detected far from the Galactic plane. A significant number of false positives is located close to the Galactic disk (about 33 per cent). Hence, removing the area of the Galaxy disk and bulge would decrease the number of false positives and objects to vet without losing (many) real transients (about 50 from $\sim$480)}
\label{fig:gallb}
\end{figure}
\begin{figure}
\includegraphics[width=\columnwidth]{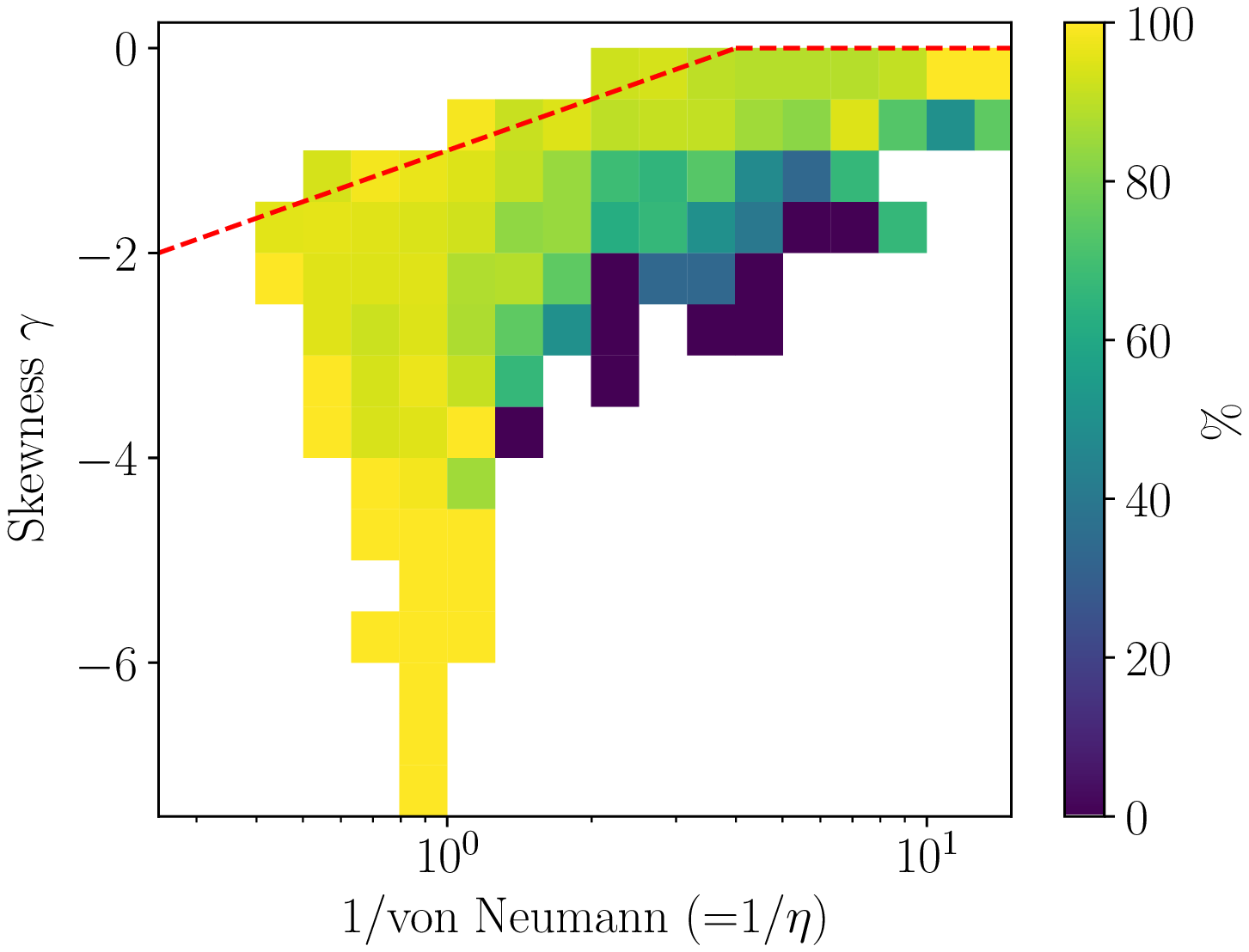}
\caption{The false-positive rate on the skewness vs. the reciprocal of von Neumann statistic plane. The percentage of rejected sources during the final vetting of the light curves and finding charts is presented. Only 8 per cent of selected objects on the skewness -- von Neumann parameter space were deemed to be transient events. There are regions on the skewness vs. the reciprocal of von Neumann statistic plane that are very efficient with finding nuclear transients candidates.}
\label{fig:vns_success}
\end{figure}

Limiting the number of false positives seems to be a crucial requirement before one can consider including a new algorithm into AlertPipe. As using the von Neumann and skewness statistics needs a significant amount of time spent on eyeballing we explored the properties of the objects classified as false positives. About 33 per cent of these objects are (manually vetted) unresolved binary systems from which 74 per cent are located close to Milky Way's disc ($|b|<25^{o}$) where only 10 per cent of transient candidates were found (see Fig. \ref{fig:gallb}). Hence, removing this area of the sky increases the number of real transients. We also noticed that using more conservative cuts on the skewness -- the reciprocal of von Neumann plane might help here (see Fig. \ref{fig:vns_success}). 
Furthermore, one can also repeat this study every $\sim$2 weeks, and announce the candidates after vetting, given that most of these transients last months to years.

\begin{figure}
\includegraphics[width=\columnwidth]{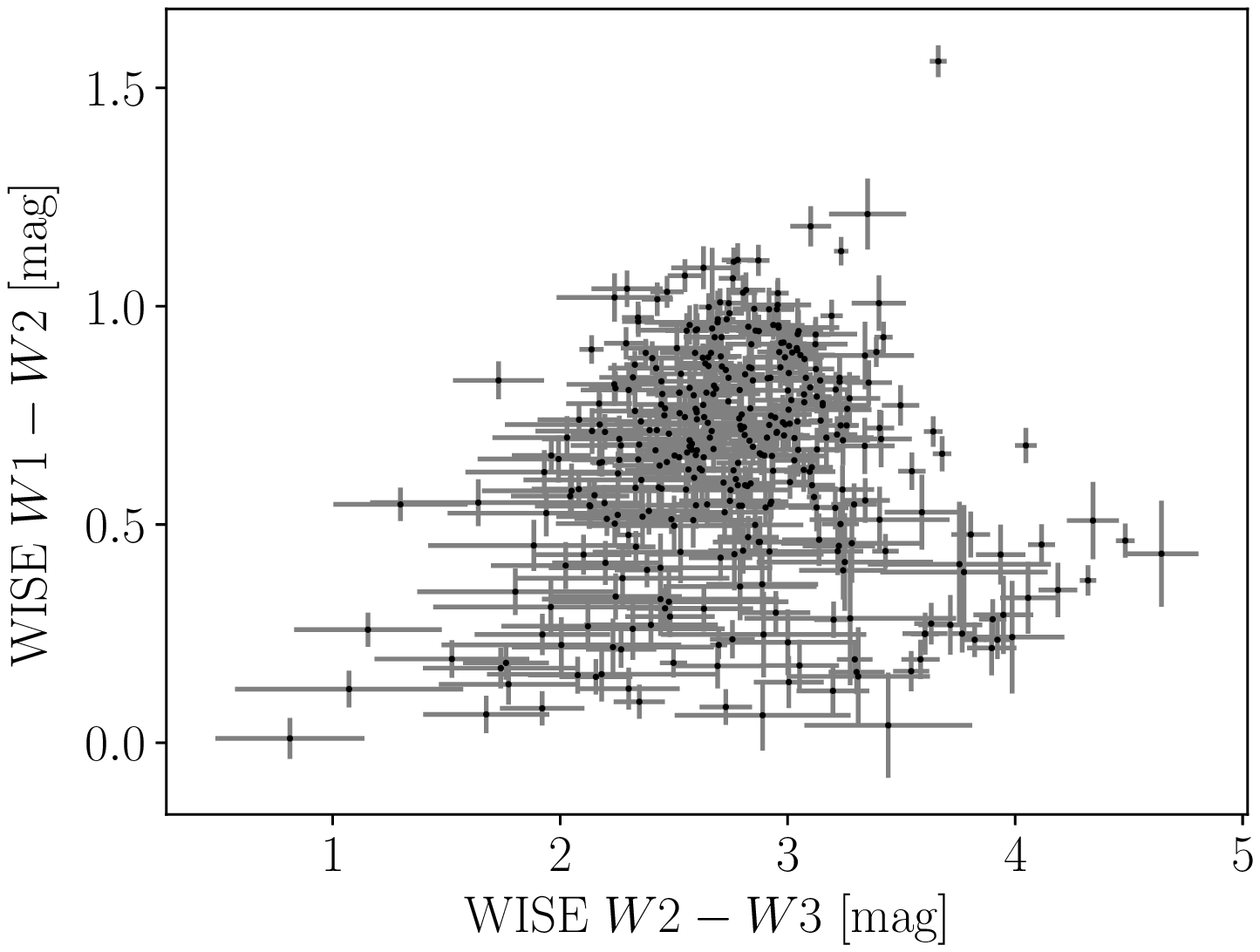}
\caption{The WISE colour-colour diagram of our sample of transients discovered in the nuclei of galaxies. For 99.6 per cent of selected objects we found a corresponding source in the AllWise data base exists. We plotted the 78 per cent of objects with robust measurements in the $W1, W2$, and $W3$ filters (i.e. the source is detected in all bands $W1, W2$, and $W3$ with a flux signal-to-noise ratio greater than 2). The sample is dominated by QSO-like objects ($W1-W2>0.5$ for $\sim$74 per cent of hosts). A small fraction of our sources have WISE colours consistent with those of elliptical galaxies ($W2-W3<\sim1$). The sample also contains spirals and starburst galaxies ($W1-W2<\sim0.5$ and $W2-W3>\sim1$). The typical colours and location on the WISE colour-colour diagram for various types of objects can be found in \citet{2010AJ....140.1868W}.}
\label{fig:wise}
\end{figure}

%http://wise2.ipac.caltech.edu/docs/release/allwise/expsup/sec2_1a.html#ph_qual
In attempt to address the nature of the transient sources discovered in our search we investigated mid-IR Wide-field Infrared Survey Explorer (WISE) data. For 99.6 per cent of our sources we obtained a cross-match within 6 arcsec (approximately the FWHM of the $W1$ WISE data) of which 78 per cent has robust measurements in the three $W1, W2$, and $W3$ filters (i.e. the detection in all bands $W1, W2$, and $W3$ has a flux signal-to-noise ratio greater than 2).
Comparing the WISE colour-colour diagram ($W2-W3$ vs.~$W1-W2$, see Fig.~\ref{fig:wise}) with that in \cite{2010AJ....140.1868W} we deduce that the majority of our sample of detected transients are from QSO-like objects.

\begin{figure}
\includegraphics[width=\columnwidth]{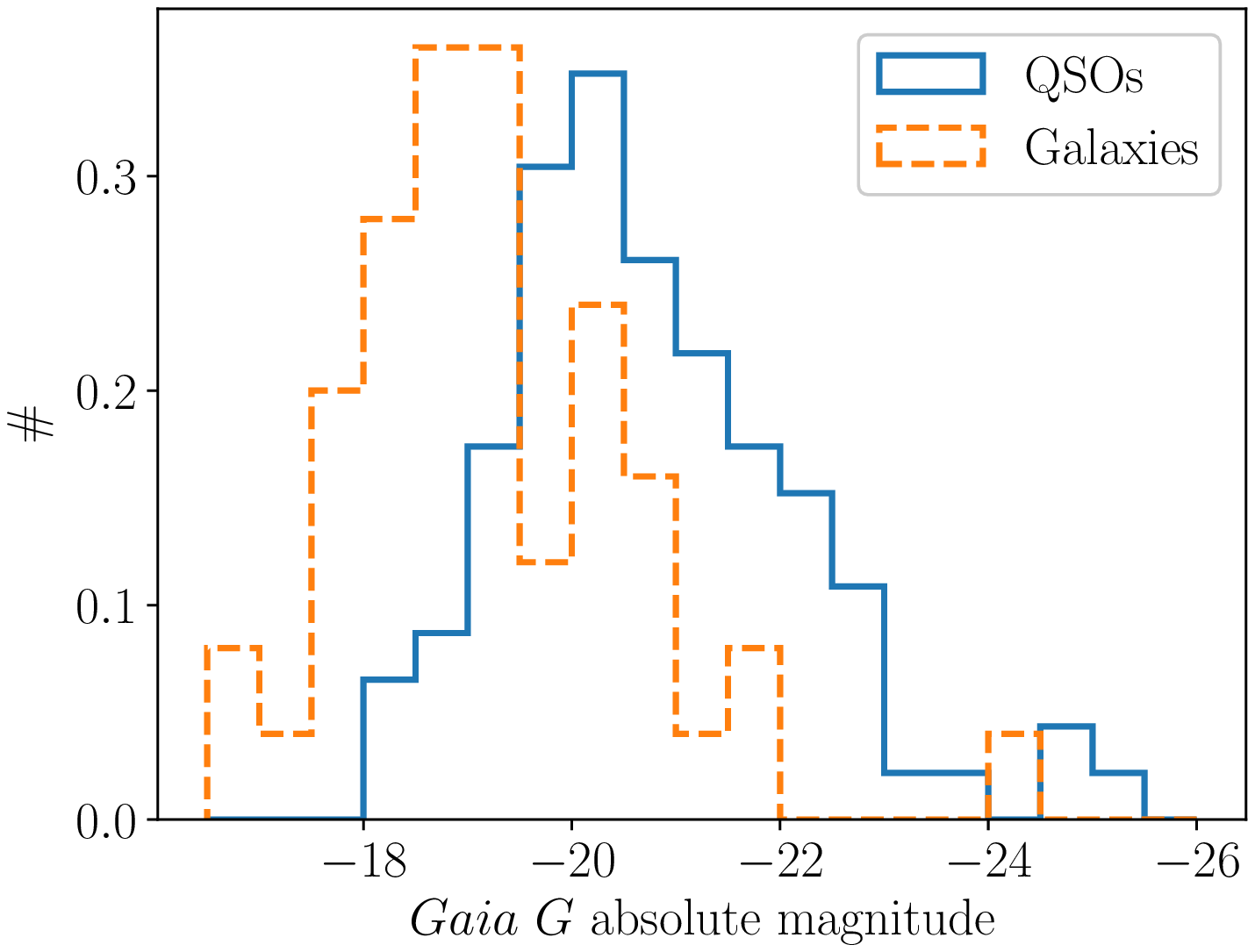}
\caption{The histogram of absolute magnitude values at the {\it Gaia} light curve peak for a subsample of the {\it Gaia} nuclear transient candidates where spectroscopic redshifts from SDSS were available. The subsample was separated into galaxies (orange dashed line) and quasars (blue line) using classification provided by SDSS. The magnitudes are host subtracted using the median value of the data from the first year of the mission (mid-2014 to mid-2015).}
\label{fig:absmag}
\end{figure}

For a subsample of the transient sources discovered in our search with spectroscopic redshifts provided by SDSS we obtained absolute magnitude values at the light curve peak. The redshift range spans between 0 and 0.6. The histogram in Figure \ref{fig:absmag} shows this subsample separated according to the SDSS classification of the source before the occurrence of the transient events, into galaxies and quasars. The absolute magnitude range spans between -17 and -26 mag with the brightest transient occur in hosts associated with quasars where the absolute magnitude of -26 mag is not unusual for this redshift range (\citealt{2018A&A...613A..51P}). Figure \ref{fig:absmag} includes 50 spectroscopically classified galaxies, and 92 QSOs, supporting our finding from the WISE data that a significant fraction of the transients we detect come from QSO-like objects. However, about one third of our nuclear transients are associated with galaxies which are not classified as AGN by SDSS. Transients detected in these galaxies could possibly arise from circumnuclear supernovae, TDEs, or AGN switching on, after they were classified in SDSS.

\subsection{Validation of {\it Gaia} Science Alerts}

During the period of 1 year between July 2016 and June 2017 GSA discovered 48 transients in galaxy nuclei (i.e. transients observed within 0.5 arcsec from their host centre if the host is recognised using external catalogues). From this sample 22 events were detected by the {\it OldSource} detector. The rest of 26 transients detected by the {\it NewSource} detector could not be found by the method described here due to the lack of the historical measurements in the light curves. Sixteen events (from the sample detected in the {\it OldSource} detector) were discovered in the SDSS objects but only 5 of them are photometrically classified as galaxies. 

We re-discovered five transients in the centres of SDSS galaxies that were previously announced as transients by the GSA team (Gaia16ajq, Gaia17ays, Gaia17bib, Gaia17cff, Gaia17dko, see Tab. \ref{tab:alrt}). One source, Gaia16avf, that was alerted on was not re-discovered by our search. Another source, Gaia17arg, was discovered on the skewness -- von Neumann plane but removed from the final list as the second field of view was pointed on the Galactic plane during the peak. One transient in the centre of an SDSS galaxy was also not found (Gaia17bje). However, the host galaxy is fainter than 20 mag in SDSS $r$--band, hence it was not included in our search.
Fifteen transients from our final sample were found by AlertPipe but then rejected through automated filtering and human visual inspection, and finally not published (see Tab. \ref{tab:unpub}). Further examination of the reasons for these rejections will be discussed in Hodgkin et al. (in prep.).

\begin{table}
\centering
\caption{The {\it Gaia} Nuclear Transient (GNT) candidates detected by GSA AlertPipe but not published due to subsequent filtering.}
\begin{tabular}{l l}
\hline
GNT ID & SDSS galaxy \\
\hline
GNTJ003643.62$+$330622.42 & SDSSJ003643.61$+$330622.52 \\ 
GNTJ042910.72$-$052040.28 & SDSSJ042910.73$-$052040.25 \\
GNTJ073442.35$+$453623.24 & SDSSJ073442.36$+$453623.28 \\
GNTJ080115.97$+$110156.53 & SDSSJ080115.97$+$110156.52 \\
GNTJ081152.11$+$252521.39 & SDSSJ081152.11$+$252521.34 \\
GNTJ143701.50$+$264019.19 & SDSSJ143701.50$+$264019.18 \\
GNTJ150512.77$+$202240.70 & SDSSJ150512.78$+$202240.70 \\
GNTJ170356.27$+$231426.67 & SDSSJ170356.27$+$231426.62 \\
GNTJ171558.78$+$362323.05 & SDSSJ171558.79$+$362323.05 \\
GNTJ172027.48$+$103210.17 & SDSSJ172027.49$+$103210.17 \\
GNTJ210213.94$+$001327.17 & SDSSJ210213.93$+$001327.18 \\
GNTJ220801.33$+$304627.97 & SDSSJ220801.33$+$304628.04 \\
GNTJ232841.41$+$224847.96 & SDSSJ232841.40$+$224848.02 \\ 
GNTJ233520.51$+$280204.32 & SDSSJ233520.51$+$280204.25 \\  
GNTJ233855.86$+$433916.86 & SDSSJ233855.86$+$433916.87 \\ 
\hline
\end{tabular}
\label{tab:unpub}
\end{table}

\begin{figure}
\includegraphics[width=\columnwidth]{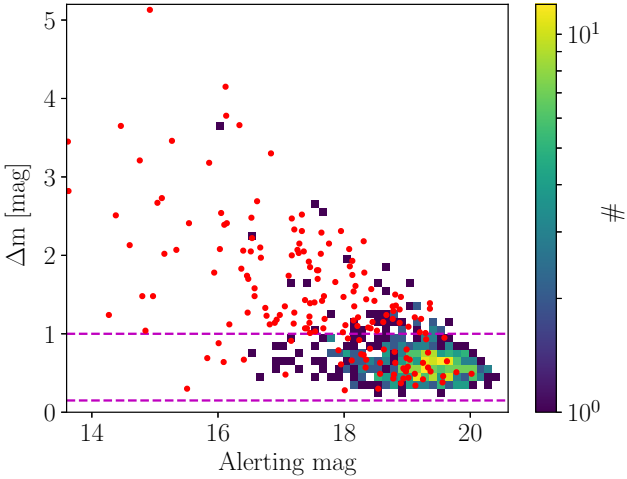}
\caption{Alerting magnitude vs. amplitude. Red dots indicate all transients (not only nuclear events) alerted by the {\it OldSource} detector in AlertPipe between July 2016 and June 2017. The squares in the background show the distribution from the sample found using the skewness -- von Neumann parameter space. The dashed magenta lines indicate the AlertPipe thresholds in the {\it OldSource} detector of delta magnitude of 0.3 and 1.0 magnitude.}
\label{fig:deltamAP}
\end{figure}

%why Gaia Alerts missed them
Figure \ref{fig:deltamAP} shows the comparison between the properties of transients found by the {\it OldSource} detector and by the search in this study. The {\it OldSource} detector and GSA filtering tends to only find the brighter transients with high amplitude whereas the transients detected on the skewness -- von Neumann parameter space are usually fainter and with lower amplitudes. 

A significant number of transients were not alerted on by the regular GSA system. There are various reasons for this situation. Sometimes multiple source IDs are assigned by the {\it Gaia} Initial Data Treatment to galaxy cores (see Subsection \ref{sec:addch} where we explain in detail how this works and how we corrected for this). We combine the magnitude measurement of different source IDs that actually belong to the same source, thereby recovering data points for the light curve of that object that increases the sensitivity of our detector. AlertPipe assumes that the {\it Gaia} Initial Data Treatment that matches sources detected during new observations with sources  those previously detected on the basis of the first pass astrometric parameters of the objects works flawlessly. This helps with removing most of the close binary systems but because the centres of galaxies are not described well by a simple PSF profile more than one source might be assigned to a galaxy core. The presence of multiple entries for the same galaxy core in the GSA DB causes GSA to exclude these events from the Alerts stream. About 45 per cent of transients found in the study presented here were flagged during {\it Gaia} detection and cross-matching as confused with different sources within the GSA DB and thus discarded by the GSA.
Another reason for a non-detection by the GSA system is that it currently requires a candidate transient to be detected at least once in each field of view within 40 days of each other, whereas {\it Gaia}'s scanning law implies that, like iPTF16fnl, the sky area of some transient sources is covered only by one of the two field of views in this 40-day window. For the sample presented in this paper, about 11 per cent of objects have a single detection within 40 days of the maximum of the {\it Gaia} light curve. The requirement of multiple detections mainly affects short transients as the second observation may happen when the source is back to quiescence and it is not the main cause for missing new transients by GSA (the {\it Gaia} scanning law was taken into account in simulations by \citealt{2016MNRAS.455..603B}). Moreover, 25 per cent of objects were detected more than once within the 40-day window, however the detection was each time in the same field of view meaning that these candidates are rejected by GSA. 
Because of these three reasons at least 56 per cent of the sources found by the method detailed in this paper could not be detected by AlertPipe. The fractions given above (45, 11, 25 per cent) cannot be directly added as a particular candidate transient might be rejected for multiple reasons. 
Several other AlertPipe thresholds set to reduce the number of false positives also reduce the number of nuclear transients such as the minimum difference between the magnitude of the latest photometric data point and the median from the previous detections. Similarly, the threshold that measures the difference between the historic variability (expressed in the rms of the light curve) and the significance of the latest data point is set such that many nuclear transients are missed as variability may be induced artificially due to the different angles with which {\it Gaia} scans over a galaxy and the observational windows of a rectangular shape (\citealt{2014sf2a.conf..421D}). 

%astrometry from Gaia 
\subsection{Future improvements}
In this paper we have demonstrated that the skewness -- von Neumann parameter space provides a new window into the discovery of transients with {\it Gaia}, which could be implemented in an improved version of AlertPipe. This naturally bypasses the existing requirement on having two fields-of-view, however, AlertPipe would need to be able to handle the significant number of {\it Gaia} sources which end up with split source IDs, which is non trivial for the current database design.

The source astrometry is obtained from a first pass of the On-Ground Attitude determination (OGA1) during the Initial Data Treatment. Using the more accurate astrometry from the second iteration (OGA2) from subsequent data processing will likely provide a boost to the study of transients in galaxy nuclei. The accuracy of the {\it Gaia} coordinates will have improved by 1 to 2 orders of magnitude and this will allow us to determine the offset between the transient and its host nucleus. However, the position of both (transient and host galaxy) must be delivered by {\it Gaia} that makes this relevant to the events detected by the {\it OldSource} detector. This is especially true if the host is present in {\it Gaia} DR2. 
We notice that several candidates for nuclear transients were rejected due to {\it Gaia} internal cross-match issues, but this should be solved after publishing {\it Gaia} Data Release 2 where the majority of close binary systems should be resolved.

\section{Summary}
We present results from an independent search for transients occurring in the centres of galaxies within the GSA DB. The search was performed using the same database, although with tools that are separate from the AlertPipe system that has been used by the GSA team to search and report transients on a daily basis. A clean sample of remaining photometrically classified galaxies from SDSS DR12 were cross-matched with the GSA DB and light curves were built from the {\it Gaia} photometry. Using mainly von Neumann and skewness statistics for light curves about 6k candidates for nuclear transients were found and manually filtered to produce a final set of $\sim$160 ($\sim$480) candidates for transients brighter than 19 mag (20.5 mag) during the period of 12 months between mid-2016 and mid-2017. However, significant manual vetting of candidate nuclear transients has been necessary to arrive to this number of 482 candidate nuclear transients as our statistical search provided about 12 times as many candidates. The sample may contain contaminants (i.e.~data artefacts and unresolved binaries). According to the WISE colours, and SDSS classification spectroscopy, a significant fraction of the transients we discovered may be due to AGN activity. Discrimination between different classes of transients and nuclear activity is impossible without extensive spectroscopic follow-up. Here, we obtained classification spectra for three candidates with peculiar transient behaviour. Implementing the even higher accuracy astrometry afforded by {\it Gaia} will be useful to confirm the nuclear nature of a transient and/or to derive the offset between the exact transient position and the galaxy's centre. This parameter space is essential for studies of tidal disruption events and other phenomena associated with galactic nuclei only.
   
\section*{Acknowledgements}
We would like to thank the referee for the useful comments that helped improve the manuscript.
We would like to thank G. Cannizzaro and T. van Grunsven for helping in the follow-up observations.
We would like to thank A. Brown and T. Prusti for reading an early version of this manuscript.
ZKR acknowledges discussions with Nadia Blagorodnova. 
ZKR and PGJ acknowledge support from European Research Council Consolidator Grant 647208. 
{\L}W acknowledges OPUS 2015/17/B/ST9/03167 grant. 
MF acknowledges support from a Royal Society Science Foundation Ireland University Research Fellowship.

This work has made use of data from the European Space Agency (ESA)
mission {\it Gaia} (\url{https://www.cosmos.esa.int/gaia}), processed by
the {\it Gaia} Data Processing and Analysis Consortium (DPAC,
\url{https://www.cosmos.esa.int/web/gaia/dpac/consortium}). Funding
for the DPAC has been provided by national institutions, in particular
the institutions participating in the {\it Gaia} Multilateral Agreement.

The ISIS/ACAM spectroscopy was obtained with the William Herschel Telescope, operated on the island of La Palma by the Isaac Newton Group of Telescopes in the Spanish Observatorio del Roque de los Muchachos of the Instituto de Astrofisica de Canarias. 

This publication has made use of data products from the Wide-field Infrared Survey Explorer, which is a joint project of the University of California, Los Angeles, and the Jet Propulsion Laboratory/California Institute of Technology, funded by the National Aeronautics and Space Administration. 

Funding for the Sloan Digital Sky Survey IV has been provided by the Alfred P. Sloan Foundation, the U.S. Department of Energy Office of Science, and the Participating Institutions. SDSS-IV acknowledges
support and resources from the Center for High-Performance Computing at
the University of Utah. The SDSS web site is www.sdss.org.
SDSS-IV is managed by the Astrophysical Research Consortium for the 
Participating Institutions of the SDSS Collaboration including the 
Brazilian Participation Group, the Carnegie Institution for Science, 
Carnegie Mellon University, the Chilean Participation Group, the French Participation Group, Harvard-Smithsonian Center for Astrophysics, 
Instituto de Astrof\'isica de Canarias, The Johns Hopkins University, 
Kavli Institute for the Physics and Mathematics of the Universe (IPMU) / 
University of Tokyo, the Korean Participation Group, Lawrence Berkeley National Laboratory, 
Leibniz Institut f\"ur Astrophysik Potsdam (AIP),  
Max-Planck-Institut f\"ur Astronomie (MPIA Heidelberg), 
Max-Planck-Institut f\"ur Astrophysik (MPA Garching), 
Max-Planck-Institut f\"ur Extraterrestrische Physik (MPE), 
National Astronomical Observatories of China, New Mexico State University, 
New York University, University of Notre Dame, 
Observat\'ario Nacional / MCTI, The Ohio State University, 
Pennsylvania State University, Shanghai Astronomical Observatory, 
United Kingdom Participation Group,
Universidad Nacional Aut\'onoma de M\'exico, University of Arizona, 
University of Colorado Boulder, University of Oxford, University of Portsmouth, 
University of Utah, University of Virginia, University of Washington, University of Wisconsin, 
Vanderbilt University, and Yale University.

This research has made use of Astropy, a community-developed core Python package for Astronomy \citep{2013A&A...558A..33A}, astroML, a Python module for machine learning and data mining \citep{astroML}, a Python module corner \citep{corner}, and Q3C extension for PostgreSQL \citep{2006ASPC..351..735K}.
%%%%%%%%%%%%%%%%%%%%%%%%%%%%%%%%%%%%%%%%%%%%%%%%%%

%%%%%%%%%%%%%%%%%%%% REFERENCES %%%%%%%%%%%%%%%%%%

% The best way to enter references is to use BibTeX:

\bibliographystyle{mnras}

\begin{thebibliography}{}
\makeatletter
\relax
\def\mn@urlcharsother{\let\do\@makeother \do\$\do\&\do\#\do\^\do\_\do\%\do\~}
\def\mn@doi{\begingroup\mn@urlcharsother \@ifnextchar [ {\mn@doi@}
  {\mn@doi@[]}}
\def\mn@doi@[#1]#2{\def\@tempa{#1}\ifx\@tempa\@empty \href
  {http://dx.doi.org/#2} {doi:#2}\else \href {http://dx.doi.org/#2} {#1}\fi
  \endgroup}
\def\mn@eprint#1#2{\mn@eprint@#1:#2::\@nil}
\def\mn@eprint@arXiv#1{\href {http://arxiv.org/abs/#1} {{\tt arXiv:#1}}}
\def\mn@eprint@dblp#1{\href {http://dblp.uni-trier.de/rec/bibtex/#1.xml}
  {dblp:#1}}
\def\mn@eprint@#1:#2:#3:#4\@nil{\def\@tempa {#1}\def\@tempb {#2}\def\@tempc
  {#3}\ifx \@tempc \@empty \let \@tempc \@tempb \let \@tempb \@tempa \fi \ifx
  \@tempb \@empty \def\@tempb {arXiv}\fi \@ifundefined
  {mn@eprint@\@tempb}{\@tempb:\@tempc}{\expandafter \expandafter \csname
  mn@eprint@\@tempb\endcsname \expandafter{\@tempc}}}

\bibitem[\protect\citeauthoryear{{Alam} et~al.,}{{Alam}
  et~al.}{2015}]{2015ApJS..219...12A}
{Alam} S.,  et~al., 2015, \mn@doi [\apjs] {10.1088/0067-0049/219/1/12}, \href
  {http://adsabs.harvard.edu/abs/2015ApJS..219...12A} {219, 12}

\bibitem[\protect\citeauthoryear{{Altavilla}, {Botticella}, {Cappellaro}  \&
  {Turatto}}{{Altavilla} et~al.}{2012}]{2012Ap&SS.341..163A}
{Altavilla} G.,  {Botticella} M.~T.,  {Cappellaro} E.,   {Turatto} M.,  2012,
  \mn@doi [\apss] {10.1007/s10509-012-1017-6}, \href
  {http://adsabs.harvard.edu/abs/2012Ap%26SS.341..163A} {341, 163}

\bibitem[\protect\citeauthoryear{{Arcavi} et~al.,}{{Arcavi}
  et~al.}{2014}]{2014ApJ...793...38A}
{Arcavi} I.,  et~al., 2014, \mn@doi [\apj] {10.1088/0004-637X/793/1/38}, \href
  {http://adsabs.harvard.edu/abs/2014ApJ...793...38A} {793, 38}

\bibitem[\protect\citeauthoryear{{Arenou} et~al.,}{{Arenou}
  et~al.}{2017}]{2017A&A...599A..50A}
{Arenou} F.,  et~al., 2017, \mn@doi [\aap] {10.1051/0004-6361/201629895}, \href
  {http://adsabs.harvard.edu/abs/2017A%26A...599A..50A} {599, A50}

\bibitem[\protect\citeauthoryear{{Astropy Collaboration} et~al.,}{{Astropy
  Collaboration} et~al.}{2013}]{2013A&A...558A..33A}
{Astropy Collaboration} et~al., 2013, \mn@doi [\aap]
  {10.1051/0004-6361/201322068}, \href
  {http://adsabs.harvard.edu/abs/2013A%26A...558A..33A} {558, A33}

\bibitem[\protect\citeauthoryear{{Belokurov} \& {Evans}}{{Belokurov} \&
  {Evans}}{2002}]{2002MNRAS.331..649B}
{Belokurov} V.~A.,  {Evans} N.~W.,  2002, \mn@doi [\mnras]
  {10.1046/j.1365-8711.2002.05222.x}, \href
  {http://adsabs.harvard.edu/abs/2002MNRAS.331..649B} {331, 649}

\bibitem[\protect\citeauthoryear{{Belokurov} \& {Evans}}{{Belokurov} \&
  {Evans}}{2003}]{2003MNRAS.341..569B}
{Belokurov} V.~A.,  {Evans} N.~W.,  2003, \mn@doi [\mnras]
  {10.1046/j.1365-8711.2003.06427.x}, \href
  {http://adsabs.harvard.edu/abs/2003MNRAS.341..569B} {341, 569}

\bibitem[\protect\citeauthoryear{{Blagorodnova}, {Van Velzen}, {Harrison},
  {Koposov}, {Mattila}, {Campbell}, {Walton}  \& {Wyrzykowski}}{{Blagorodnova}
  et~al.}{2016}]{2016MNRAS.455..603B}
{Blagorodnova} N.,  {Van Velzen} S.,  {Harrison} D.~L.,  {Koposov} S.,
  {Mattila} S.,  {Campbell} H.,  {Walton} N.~A.,   {Wyrzykowski} {\L}.,  2016,
  \mn@doi [\mnras] {10.1093/mnras/stv2308}, \href
  {http://adsabs.harvard.edu/abs/2016MNRAS.455..603B} {455, 603}

\bibitem[\protect\citeauthoryear{{Blagorodnova} et~al.,}{{Blagorodnova}
  et~al.}{2017}]{2017ApJ...844...46B}
{Blagorodnova} N.,  et~al., 2017, \mn@doi [\apj] {10.3847/1538-4357/aa7579},
  \href {http://adsabs.harvard.edu/abs/2017ApJ...844...46B} {844, 46}

\bibitem[\protect\citeauthoryear{{Blondin} \& {Tonry}}{{Blondin} \&
  {Tonry}}{2007}]{2007ApJ...666.1024B}
{Blondin} S.,  {Tonry} J.~L.,  2007, \mn@doi [\apj] {10.1086/520494}, \href
  {http://adsabs.harvard.edu/abs/2007ApJ...666.1024B} {666, 1024}

\bibitem[\protect\citeauthoryear{{Carrasco} et~al.,}{{Carrasco}
  et~al.}{2016}]{2016A&A...595A...7C}
{Carrasco} J.~M.,  et~al., 2016, \mn@doi [\aap] {10.1051/0004-6361/201629235},
  \href {http://adsabs.harvard.edu/abs/2016A%26A...595A...7C} {595, A7}

\bibitem[\protect\citeauthoryear{{Chen}, {Sesana}, {Madau}  \& {Liu}}{{Chen}
  et~al.}{2011}]{2011ApJ...729...13C}
{Chen} X.,  {Sesana} A.,  {Madau} P.,   {Liu} F.~K.,  2011, \mn@doi [\apj]
  {10.1088/0004-637X/729/1/13}, \href
  {http://adsabs.harvard.edu/abs/2011ApJ...729...13C} {729, 13}

\bibitem[\protect\citeauthoryear{{Dessart} et~al.,}{{Dessart}
  et~al.}{2014}]{2014MNRAS.440.1856D}
{Dessart} L.,  et~al., 2014, \mn@doi [\mnras] {10.1093/mnras/stu417}, \href
  {http://adsabs.harvard.edu/abs/2014MNRAS.440.1856D} {440, 1856}

\bibitem[\protect\citeauthoryear{{Dong} et~al.,}{{Dong}
  et~al.}{2016}]{2016Sci...351..257D}
{Dong} S.,  et~al., 2016, \mn@doi [Science] {10.1126/science.aac9613}, \href
  {http://adsabs.harvard.edu/abs/2016Sci...351..257D} {351, 257}

\bibitem[\protect\citeauthoryear{{Ducourant}, {Krone-Martins}, {Galluccio}  \&
  {Teixeira}}{{Ducourant} et~al.}{2014}]{2014sf2a.conf..421D}
{Ducourant} C.,  {Krone-Martins} A.,  {Galluccio} L.,   {Teixeira} R.,  2014,
  in {Ballet} J.,  {Martins} F.,  {Bournaud} F.,  {Monier} R.,   {Reyl{\'e}}
  C.,  eds, SF2A-2014: Proceedings of the Annual meeting of the French Society
  of Astronomy and Astrophysics. pp 421--425

\bibitem[\protect\citeauthoryear{{Eyer} et~al.,}{{Eyer}
  et~al.}{2017}]{2017arXiv170203295E}
{Eyer} L.,  et~al., 2017, preprint, \href
  {http://adsabs.harvard.edu/abs/2017arXiv170203295E} {} (\mn@eprint {arXiv}
  {1702.03295})

\bibitem[\protect\citeauthoryear{{Fabricius} et~al.,}{{Fabricius}
  et~al.}{2016}]{2016A&A...595A...3F}
{Fabricius} C.,  et~al., 2016, \mn@doi [\aap] {10.1051/0004-6361/201628643},
  \href {http://adsabs.harvard.edu/abs/2016A%26A...595A...3F} {595, A3}

\bibitem[\protect\citeauthoryear{Foreman-Mackey}{Foreman-Mackey}{2016}]{corner}
Foreman-Mackey D.,  2016, \mn@doi [The Journal of Open Source Software]
  {10.21105/joss.00024}, 24

\bibitem[\protect\citeauthoryear{{Gaia Collaboration} et~al.,}{{Gaia
  Collaboration} et~al.}{2016a}]{2016A&A...595A...1G}
{Gaia Collaboration} et~al., 2016a, \mn@doi [\aap]
  {10.1051/0004-6361/201629272}, \href
  {http://adsabs.harvard.edu/abs/2016A%26A...595A...1G} {595, A1}

\bibitem[\protect\citeauthoryear{{Gaia Collaboration} et~al.,}{{Gaia
  Collaboration} et~al.}{2016b}]{2016A&A...595A...2G}
{Gaia Collaboration} et~al., 2016b, \mn@doi [\aap]
  {10.1051/0004-6361/201629512}, \href
  {http://adsabs.harvard.edu/abs/2016A%26A...595A...2G} {595, A2}

\bibitem[\protect\citeauthoryear{{Gal-Yam}}{{Gal-Yam}}{2012}]{2012Sci...337..927G}
{Gal-Yam} A.,  2012, \mn@doi [Science] {10.1126/science.1203601}, \href
  {http://adsabs.harvard.edu/abs/2012Sci...337..927G} {337, 927}

\bibitem[\protect\citeauthoryear{{Graham}, {Djorgovski}, {Drake}, {Stern},
  {Mahabal}, {Glikman}, {Larson}  \& {Christensen}}{{Graham}
  et~al.}{2017}]{2017MNRAS.470.4112G}
{Graham} M.~J.,  {Djorgovski} S.~G.,  {Drake} A.~J.,  {Stern} D.,  {Mahabal}
  A.~A.,  {Glikman} E.,  {Larson} S.,   {Christensen} E.,  2017, \mn@doi
  [\mnras] {10.1093/mnras/stx1456}, \href
  {http://adsabs.harvard.edu/abs/2017MNRAS.470.4112G} {470, 4112}

\bibitem[\protect\citeauthoryear{{Hodgkin}, {Wyrzykowski}, {Blagorodnova}  \&
  {Koposov}}{{Hodgkin} et~al.}{2013}]{2013RSPTA.37120239H}
{Hodgkin} S.~T.,  {Wyrzykowski} L.,  {Blagorodnova} N.,   {Koposov} S.,  2013,
  \mn@doi [Philosophical Transactions of the Royal Society of London Series A]
  {10.1098/rsta.2012.0239}, \href
  {http://adsabs.harvard.edu/abs/2013RSPTA.37120239H} {371, 20120239}

\bibitem[\protect\citeauthoryear{{Holoien} et~al.,}{{Holoien}
  et~al.}{2016}]{2016MNRAS.455.2918H}
{Holoien} T.~W.-S.,  et~al., 2016, \mn@doi [\mnras] {10.1093/mnras/stv2486},
  \href {http://adsabs.harvard.edu/abs/2016MNRAS.455.2918H} {455, 2918}

\bibitem[\protect\citeauthoryear{{Inserra} et~al.,}{{Inserra}
  et~al.}{2018}]{2018MNRAS.475.1046I}
{Inserra} C.,  et~al., 2018, \mn@doi [\mnras] {10.1093/mnras/stx3179}, \href
  {http://adsabs.harvard.edu/abs/2018MNRAS.475.1046I} {475, 1046}

\bibitem[\protect\citeauthoryear{{Ivezi{\'c}}, {Connelly}, {VanderPlas}  \&
  {Gray}}{{Ivezi{\'c}} et~al.}{2014}]{2014sdmm.book.....I}
{Ivezi{\'c}} {\v Z}.,  {Connelly} A.~J.,  {VanderPlas} J.~T.,   {Gray} A.,
  2014, {Statistics, Data Mining, and Machine Learningin Astronomy}.
Princeton University Press

\bibitem[\protect\citeauthoryear{{Kankare} et~al.,}{{Kankare}
  et~al.}{2017}]{2017NatAs...1..865K}
{Kankare} E.,  et~al., 2017, \mn@doi [Nature Astronomy]
  {10.1038/s41550-017-0290-2}, \href
  {http://adsabs.harvard.edu/abs/2017NatAs...1..865K} {1, 865}

\bibitem[\protect\citeauthoryear{{Kochanek}}{{Kochanek}}{2016}]{2016MNRAS.461..371K}
{Kochanek} C.~S.,  2016, \mn@doi [\mnras] {10.1093/mnras/stw1290}, \href
  {http://adsabs.harvard.edu/abs/2016MNRAS.461..371K} {461, 371}

\bibitem[\protect\citeauthoryear{{Koposov} \& {Bartunov}}{{Koposov} \&
  {Bartunov}}{2006}]{2006ASPC..351..735K}
{Koposov} S.,  {Bartunov} O.,  2006, in {Gabriel} C.,  {Arviset} C.,  {Ponz}
  D.,   {Enrique} S.,  eds,  Astronomical Society of the Pacific Conference
  Series Vol. 351, Astronomical Data Analysis Software and Systems XV. p.~735

\bibitem[\protect\citeauthoryear{{Koz{\l}owski} et~al.,}{{Koz{\l}owski}
  et~al.}{2010}]{2010ApJ...708..927K}
{Koz{\l}owski} S.,  et~al., 2010, \mn@doi [\apj] {10.1088/0004-637X/708/2/927},
  \href {http://adsabs.harvard.edu/abs/2010ApJ...708..927K} {708, 927}

\bibitem[\protect\citeauthoryear{{Leloudas} et~al.,}{{Leloudas}
  et~al.}{2016}]{2016NatAs...1E...2L}
{Leloudas} G.,  et~al., 2016, \mn@doi [Nature Astronomy]
  {10.1038/s41550-016-0002}, \href
  {http://adsabs.harvard.edu/abs/2016NatAs...1E...2L} {1, 0002}

\bibitem[\protect\citeauthoryear{{Lindegren} et~al.,}{{Lindegren}
  et~al.}{2016}]{2016A&A...595A...4L}
{Lindegren} L.,  et~al., 2016, \mn@doi [\aap] {10.1051/0004-6361/201628714},
  \href {http://adsabs.harvard.edu/abs/2016A%26A...595A...4L} {595, A4}

\bibitem[\protect\citeauthoryear{{MacLeod} et~al.,}{{MacLeod}
  et~al.}{2010}]{2010ApJ...721.1014M}
{MacLeod} C.~L.,  et~al., 2010, \mn@doi [\apj] {10.1088/0004-637X/721/2/1014},
  \href {http://adsabs.harvard.edu/abs/2010ApJ...721.1014M} {721, 1014}

\bibitem[\protect\citeauthoryear{{MacLeod} et~al.,}{{MacLeod}
  et~al.}{2012}]{2012ApJ...753..106M}
{MacLeod} C.~L.,  et~al., 2012, \mn@doi [\apj] {10.1088/0004-637X/753/2/106},
  \href {http://adsabs.harvard.edu/abs/2012ApJ...753..106M} {753, 106}

\bibitem[\protect\citeauthoryear{{Maoz} \& {Graur}}{{Maoz} \&
  {Graur}}{2017}]{2017ApJ...848...25M}
{Maoz} D.,  {Graur} O.,  2017, \mn@doi [\apj] {10.3847/1538-4357/aa8b6e}, \href
  {http://adsabs.harvard.edu/abs/2017ApJ...848...25M} {848, 25}

\bibitem[\protect\citeauthoryear{{Newberg}, {Richards}, {Richmond}  \&
  {Fan}}{{Newberg} et~al.}{1999}]{1999ApJS..123..377N}
{Newberg} H.~J.,  {Richards} G.~T.,  {Richmond} M.,   {Fan} X.,  1999, \mn@doi
  [\apjs] {10.1086/313241}, \href
  {http://adsabs.harvard.edu/abs/1999ApJS..123..377N} {123, 377}

\bibitem[\protect\citeauthoryear{{Nugent}, {Kim}  \& {Perlmutter}}{{Nugent}
  et~al.}{2002}]{2002PASP..114..803N}
{Nugent} P.,  {Kim} A.,   {Perlmutter} S.,  2002, \mn@doi [\pasp]
  {10.1086/341707}, \href {http://adsabs.harvard.edu/abs/2002PASP..114..803N}
  {114, 803}

\bibitem[\protect\citeauthoryear{{P{\^a}ris} et~al.,}{{P{\^a}ris}
  et~al.}{2018}]{2018A&A...613A..51P}
{P{\^a}ris} I.,  et~al., 2018, \mn@doi [\aap] {10.1051/0004-6361/201732445},
  \href {http://adsabs.harvard.edu/abs/2018A%26A...613A..51P} {613, A51}

\bibitem[\protect\citeauthoryear{{Pastorello} et~al.,}{{Pastorello}
  et~al.}{2010}]{2010ApJ...724L..16P}
{Pastorello} A.,  et~al., 2010, \mn@doi [\apjl] {10.1088/2041-8205/724/1/L16},
  \href {http://adsabs.harvard.edu/abs/2010ApJ...724L..16P} {724, L16}

\bibitem[\protect\citeauthoryear{{Price-Whelan} et~al.,}{{Price-Whelan}
  et~al.}{2014}]{2014ApJ...781...35P}
{Price-Whelan} A.~M.,  et~al., 2014, \mn@doi [\apj]
  {10.1088/0004-637X/781/1/35}, \href
  {http://adsabs.harvard.edu/abs/2014ApJ...781...35P} {781, 35}

\bibitem[\protect\citeauthoryear{{Rattenbury} et~al.,}{{Rattenbury}
  et~al.}{2015}]{2015MNRAS.447L..31R}
{Rattenbury} N.~J.,  et~al., 2015, \mn@doi [\mnras] {10.1093/mnrasl/slu176},
  \href {http://adsabs.harvard.edu/abs/2015MNRAS.447L..31R} {447, L31}

\bibitem[\protect\citeauthoryear{{Rees}}{{Rees}}{1988}]{1988Natur.333..523R}
{Rees} M.~J.,  1988, \mn@doi [\nat] {10.1038/333523a0}, \href
  {http://adsabs.harvard.edu/abs/1988Natur.333..523R} {333, 523}

\bibitem[\protect\citeauthoryear{{Stone} \& {Metzger}}{{Stone} \&
  {Metzger}}{2016}]{2016MNRAS.455..859S}
{Stone} N.~C.,  {Metzger} B.~D.,  2016, \mn@doi [\mnras]
  {10.1093/mnras/stv2281}, \href
  {http://adsabs.harvard.edu/abs/2016MNRAS.455..859S} {455, 859}

\bibitem[\protect\citeauthoryear{{Stoughton} et~al.,}{{Stoughton}
  et~al.}{2002}]{2002AJ....123..485S}
{Stoughton} C.,  et~al., 2002, \mn@doi [\aj] {10.1086/324741}, \href
  {http://adsabs.harvard.edu/abs/2002AJ....123..485S} {123, 485}

\bibitem[\protect\citeauthoryear{{Strateva} et~al.,}{{Strateva}
  et~al.}{2001}]{2001AJ....122.1861S}
{Strateva} I.,  et~al., 2001, \mn@doi [\aj] {10.1086/323301}, \href
  {http://adsabs.harvard.edu/abs/2001AJ....122.1861S} {122, 1861}

\bibitem[\protect\citeauthoryear{{Taddia} et~al.,}{{Taddia}
  et~al.}{2017}]{2017ATel10105....1T}
{Taddia} F.,  et~al., 2017, The Astronomer's Telegram, \href
  {http://adsabs.harvard.edu/abs/2017ATel10105....1T} {105}

\bibitem[\protect\citeauthoryear{{Vanderplas}, {Connolly}, {Ivezi{\'c}}  \&
  {Gray}}{{Vanderplas} et~al.}{2012}]{astroML}
{Vanderplas} J.,  {Connolly} A.,  {Ivezi{\'c}} {\v Z}.,   {Gray} A.,  2012, in
  Conference on Intelligent Data Understanding (CIDU). pp 47 --54,
  \mn@doi{10.1109/CIDU.2012.6382200}

\bibitem[\protect\citeauthoryear{{Wegg} \& {Nate Bode}}{{Wegg} \& {Nate
  Bode}}{2011}]{2011ApJ...738L...8W}
{Wegg} C.,  {Nate Bode} J.,  2011, \mn@doi [\apjl]
  {10.1088/2041-8205/738/1/L8}, \href
  {http://adsabs.harvard.edu/abs/2011ApJ...738L...8W} {738, L8}

\bibitem[\protect\citeauthoryear{{Wevers} et~al.,}{{Wevers}
  et~al.}{2018}]{2018MNRAS.473.3854W}
{Wevers} T.,  et~al., 2018, \mn@doi [\mnras] {10.1093/mnras/stx2625}, \href
  {http://adsabs.harvard.edu/abs/2018MNRAS.473.3854W} {473, 3854}

\bibitem[\protect\citeauthoryear{{Wright} et~al.,}{{Wright}
  et~al.}{2010}]{2010AJ....140.1868W}
{Wright} E.~L.,  et~al., 2010, \mn@doi [\aj] {10.1088/0004-6256/140/6/1868},
  \href {http://adsabs.harvard.edu/abs/2010AJ....140.1868W} {140, 1868}

\bibitem[\protect\citeauthoryear{{Wyrzykowski} \& {Hodgkin}}{{Wyrzykowski} \&
  {Hodgkin}}{2012}]{2012IAUS..285..425W}
{Wyrzykowski} {\L}.,  {Hodgkin} S.,  2012, in {Griffin} E.,  {Hanisch} R.,
  {Seaman} R.,  eds,  IAU Symposium Vol. 285, New Horizons in Time Domain
  Astronomy. pp 425--428 (\mn@eprint {arXiv} {1112.0187}),
  \mn@doi{10.1017/S1743921312001305}

\bibitem[\protect\citeauthoryear{{Wyrzykowski} et~al.,}{{Wyrzykowski}
  et~al.}{2016}]{2016MNRAS.458.3012W}
{Wyrzykowski} {\L}.,  et~al., 2016, \mn@doi [\mnras] {10.1093/mnras/stw426},
  \href {http://adsabs.harvard.edu/abs/2016MNRAS.458.3012W} {458, 3012}

\bibitem[\protect\citeauthoryear{{Wyrzykowski} et~al.,}{{Wyrzykowski}
  et~al.}{2017}]{2017MNRAS.465L.114W}
{Wyrzykowski} {\L}.,  et~al., 2017, \mn@doi [\mnras] {10.1093/mnrasl/slw213},
  \href {http://adsabs.harvard.edu/abs/2017MNRAS.465L.114W} {465, L114}

\bibitem[\protect\citeauthoryear{{Zackay}, {Ofek}  \& {Gal-Yam}}{{Zackay}
  et~al.}{2016}]{2016ApJ...830...27Z}
{Zackay} B.,  {Ofek} E.~O.,   {Gal-Yam} A.,  2016, \mn@doi [\apj]
  {10.3847/0004-637X/830/1/27}, \href
  {http://adsabs.harvard.edu/abs/2016ApJ...830...27Z} {830, 27}

\bibitem[\protect\citeauthoryear{{de Bruijne}, {Allen}, {Azaz},
  {Krone-Martins}, {Prod'homme}  \& {Hestroffer}}{{de Bruijne}
  et~al.}{2015}]{2015A&A...576A..74D}
{de Bruijne} J.~H.~J.,  {Allen} M.,  {Azaz} S.,  {Krone-Martins} A.,
  {Prod'homme} T.,   {Hestroffer} D.,  2015, \mn@doi [\aap]
  {10.1051/0004-6361/201424018}, \href
  {http://adsabs.harvard.edu/abs/2015A%26A...576A..74D} {576, A74}

\bibitem[\protect\citeauthoryear{{van Dokkum}}{{van
  Dokkum}}{2001}]{2001PASP..113.1420V}
{van Dokkum} P.~G.,  2001, \mn@doi [\pasp] {10.1086/323894}, \href
  {http://adsabs.harvard.edu/abs/2001PASP..113.1420V} {113, 1420}

\bibitem[\protect\citeauthoryear{{van Leeuwen} et~al.,}{{van Leeuwen}
  et~al.}{2017}]{2017A&A...599A..32V}
{van Leeuwen} F.,  et~al., 2017, \mn@doi [\aap] {10.1051/0004-6361/201630064},
  \href {http://adsabs.harvard.edu/abs/2017A%26A...599A..32V} {599, A32}

\bibitem[\protect\citeauthoryear{{van Velzen}}{{van
  Velzen}}{2018}]{2018ApJ...852...72V}
{van Velzen} S.,  2018, \mn@doi [\apj] {10.3847/1538-4357/aa998e}, \href
  {http://adsabs.harvard.edu/abs/2018ApJ...852...72V} {852, 72}

\bibitem[\protect\citeauthoryear{von Neumann}{von
  Neumann}{1941}]{vonneumann1941}
von Neumann J.,  1941, \mn@doi [Ann. Math. Statist.] {10.1214/aoms/1177731677},
  12, 367

\makeatother
\end{thebibliography}

%%%%%%%%%%%%%%%%%%%%%%%%%%%%%%%%%%%%%%%%%%%%%%%%%%

%%%%%%%%%%%%%%%%% APPENDICES %%%%%%%%%%%%%%%%%%%%%

%%%%%%%%%%%%%%%%%%%%%%%%%%%%%%%%%%%%%%%%%%%%%%%%%%

% Don't change these lines
\bsp	% typesetting comment
\label{lastpage}
\end{document}